\documentclass[aps,prd,superscriptaddress,notitlepage,reprint]{revtex4-1}
\usepackage{graphicx}
\usepackage{isotope}

\begin{document}

\bibliographystyle{apsrev4-1-etal}

\title{Reactor Neutrino Spectral Distortions\\Play Little Role in Mass Hierarchy Experiments}
\preprint{LA-UR-18-23776}

\author{D. L. Danielson}
\email{dldanielson@ucdavis.edu}
\affiliation{Theoretical Division, Los Alamos National Laboratory, Los Alamos, New Mexico 87545, USA}
\affiliation{Department of Physics, University of California at Davis, Davis, California 95616, USA}
\author{A. C. Hayes}
\affiliation{Theoretical Division, Los Alamos National Laboratory, Los Alamos, New Mexico 87545, USA}
\author{G. T. Garvey}
\affiliation{Physics Division, Los Alamos National Laboratory, Los Alamos, New Mexico 87545, USA}
\affiliation{Department of Physics, University of Washington, Seattle, Washington 98195, USA}

\date{\today}
\begin{abstract}
\noindent The Coulomb enhancement of low energy electrons in nuclear beta decay generates sharp cutoffs in the accompanying antineutrino spectrum at the beta decay endpoint energies. It has been conjectured that these features will interfere with measuring the effect of a neutrino mass hierarchy on an oscillated nuclear reactor antineutrino spectrum. These sawtooth-like features will appear in detailed reactor antineutrino spectra, with characteristic energy scales similar to the oscillation period critical to neutrino mass hierarchy determination near a 53 km baseline. However, these sawtooth-like distortions are found to contribute at a magnitude of only a few percent relative to the mass hierarchy-dependent oscillation pattern in Fourier space. In the Fourier cosine and sine transforms, the features that encode a neutrino mass hierarchy dominate by over sixteen (thirty-three) times in prominence to the maximal contribution of the sawtooth-like distortions from the detailed energy spectrum, given $3.2\%/\sqrt{E_\mathrm{vis.}/\mathrm{MeV}}$ (perfect) detector energy resolution. The effect of these distortions is shown to be negligible even when the uncertainties in the reactor spectrum, oscillation parameters, and counting statistics are considered. This result is shown to hold even when the opposite hierarchy oscillation patterns are nearly degenerate in energy space, if energy response nonlinearities are controlled to below 0.5\%. Therefore with accurate knowledge of detector energy response, the sawtooth-like features in reactor antineutrino spectra will not significantly impede neutrino mass hierarchy measurements using reactor antineutrinos.
\end{abstract}

\maketitle

\newpage

\section{Introduction} \label{sec:intro}

The Jiangmen Underground Neutrino Observatory (JUNO) \cite{JUNO} aims to determine the neutrino mass hierarchy by measuring the oscillations
of reactor antineutrinos with very good ($\sim$3\%/$\sqrt{E_\mathrm{vis.}/\mathrm{MeV}}$) energy resolution, where $E_\mathrm{vis.}$ denotes the energy visible to the detector from an inverse beta decay.
In the standard three neutrino flavor picture there are two neutrino mass-squared differences: $\Delta m_{21}^2\equiv m^2_2-m^2_1\simeq 7.55\times 10^{-5} \mathrm{eV}^2$ and 
$\Delta m_{31}^2\equiv m^2_3-m^2_1 \simeq +2.50\:(-2.42)\times 10^{-3} \mathrm{eV}^2$---the sign and precise magnitude of $\Delta m_{31}^2$ being unknown.
Thus, there are two possible mass hierarchies: 
one with so-called `normal' mass ordering ($m_1<m_2\ll m_3$) and one with `inverted' mass ordering ($m_3\ll m_1<m_2$).
Any fundamental physics explanation of neutrino masses and their mixing requires knowledge of which of these two 
mass orderings appears in nature.

Measuring a neutrino mass hierarchy with neutrino oscillation experiments is difficult and generally requires either very long baseline matter-induced effects to introduce an additional phase relative to vacuum oscillations, 
or non-negligible mixing among all three flavors in vacuum.
Petcov and Piai \cite{Petcov}
 have pointed out that the oscillations of antineutrinos from nuclear reactors could, in principle, be used to
measure the mass hierarchy via the latter method, using a baseline distance of several tens of kilometers.
Even at the optimum baseline ($\sim$50~km), 
where oscillations from $\Delta m^2_{21}$ are maximally suppressed and sensitivity to $\Delta m^2_{31}$ enhanced, such a hierarchy 
measurement is a very challenging one because it requires sensitivity to the small and rapid oscillations from $\Delta m^2_{31}$ and $\Delta m^2_{32}$, 
which necessarily requires unprecedented energy resolution.

Another difficulty accompanies certain experimentally allowed oscillation parameter values, for which a wrong-hierarchy spectrum can be fitted to the true spectrum with only a few percent difference per energy bin \cite{lisi}. In such cases even small spectral uncertainties render the hierarchies degenerate in energy space, though Fourier analysis may resolve the degeneracy if these uncertainties are decoupled from the hierarchy-dependent oscillations and if energy scale nonlinearities are controlled with unprecedented accuracy within tenths of one percent \cite{qian}.

We defer to other authors' important and ongoing analyses of these experimental challenges. We also defer the consideration of large reactor-specific variations in the shape of the antineutrino spectrum, restricting our focus to the impact of the generic fine structure of reactor antineutrino spectra. The purpose of the present work is to examine the conjecture \cite{forero} that limited knowledge of the small features of these spectra can have a significant effect on mass hierarchy measurements using reactor antineutrinos.
In particular we investigate whether the sawtooth-like structures in reactor antineutrino spectra, which may align with mass hierarchy-dependent oscillations, pose a serious challenge to neutrino mass hierarchy experiments.

The remainder of this work comprises four sections. Section~\ref{sec:primary} demonstrates the very small scale of the sawtooth-like spectral distortions relative to hierarchy-dependent oscillations, with \ref{sec:beta} analyzing the fine structure of the beta decay spectrum, and \ref{sec:fourier} presenting Fourier analysis of the resulting oscillated antineutrino spectra. Section~\ref{sec:uncertainties} demonstrates that these distortions remain negligibly small throughout the uncertainty ranges on the beta decay (\ref{sec:decayUncertainties}) and oscillation (\ref{sec:osc}) parameters. Section~\ref{sec:challenges} addresses whether the fine structure of the reactor spectrum can become non-negligible when compounded with the challenges posed by nearly indistinguishable opposite-hierarchy oscillation patterns (\ref{sec:hierarchyDegeneracies}) and significant nonlinearities in detector energy response (\ref{sec:nonlinearities}). Finally, Section~\ref{sec:conclusion} presents the overall effect of the fine structure of the reactor spectrum in Fourier space (\ref{sec:zoomedOut}) and summarizes our findings (\ref{sec:summary}).

\begin{figure}
\includegraphics[width= 0.9\linewidth]{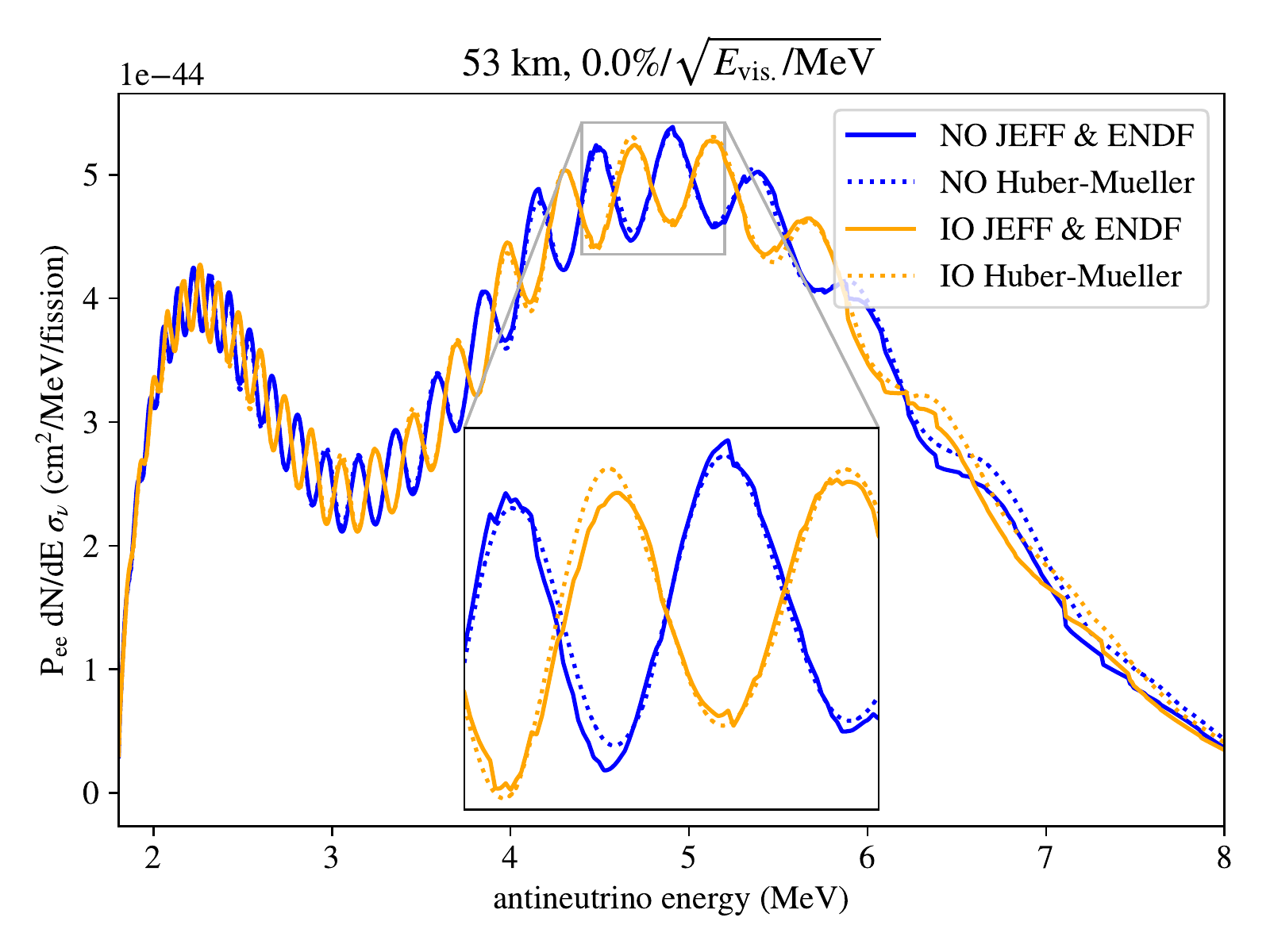}
\includegraphics[width= 0.9\linewidth]{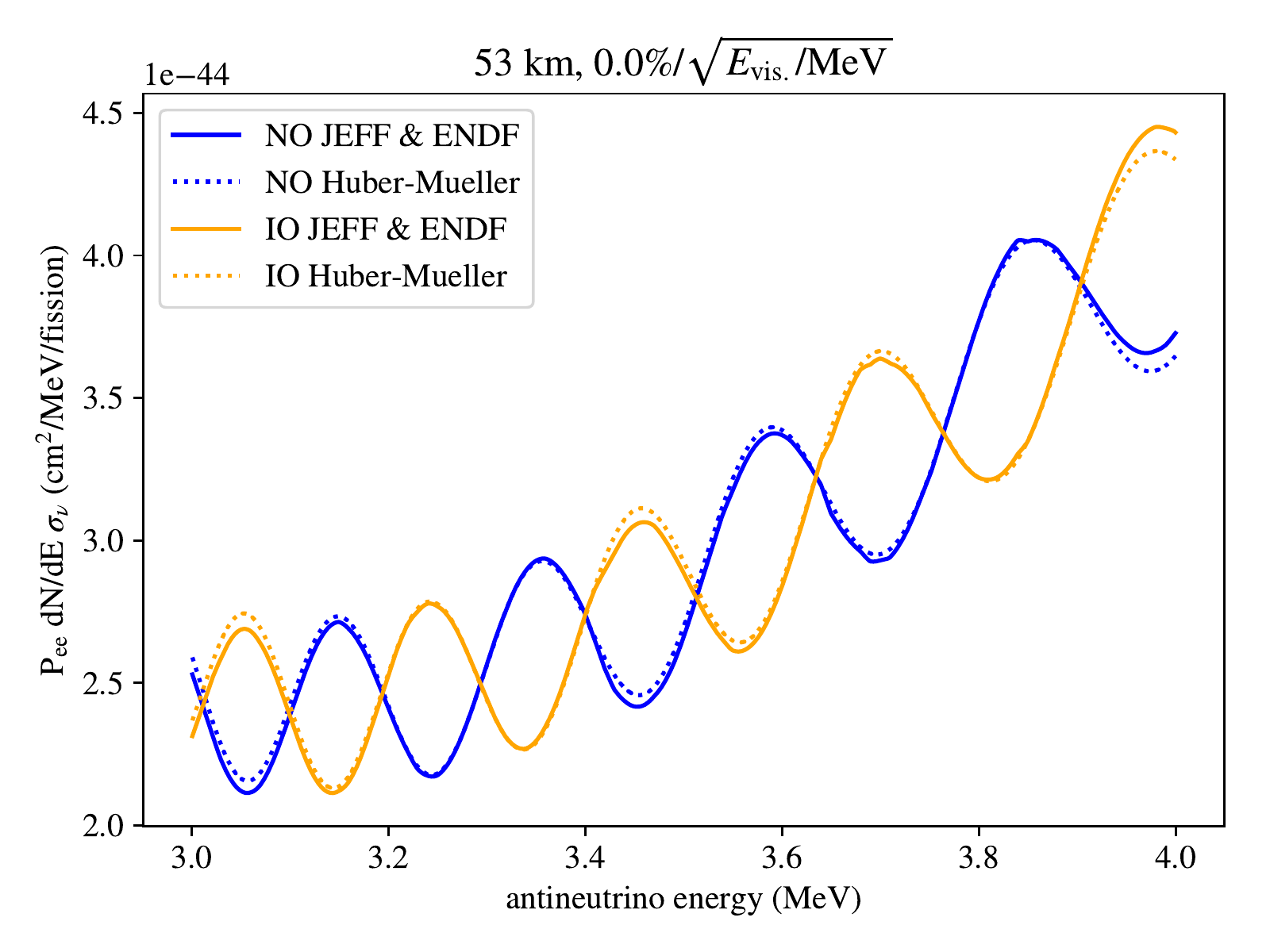}
\includegraphics[width= 0.9\linewidth]{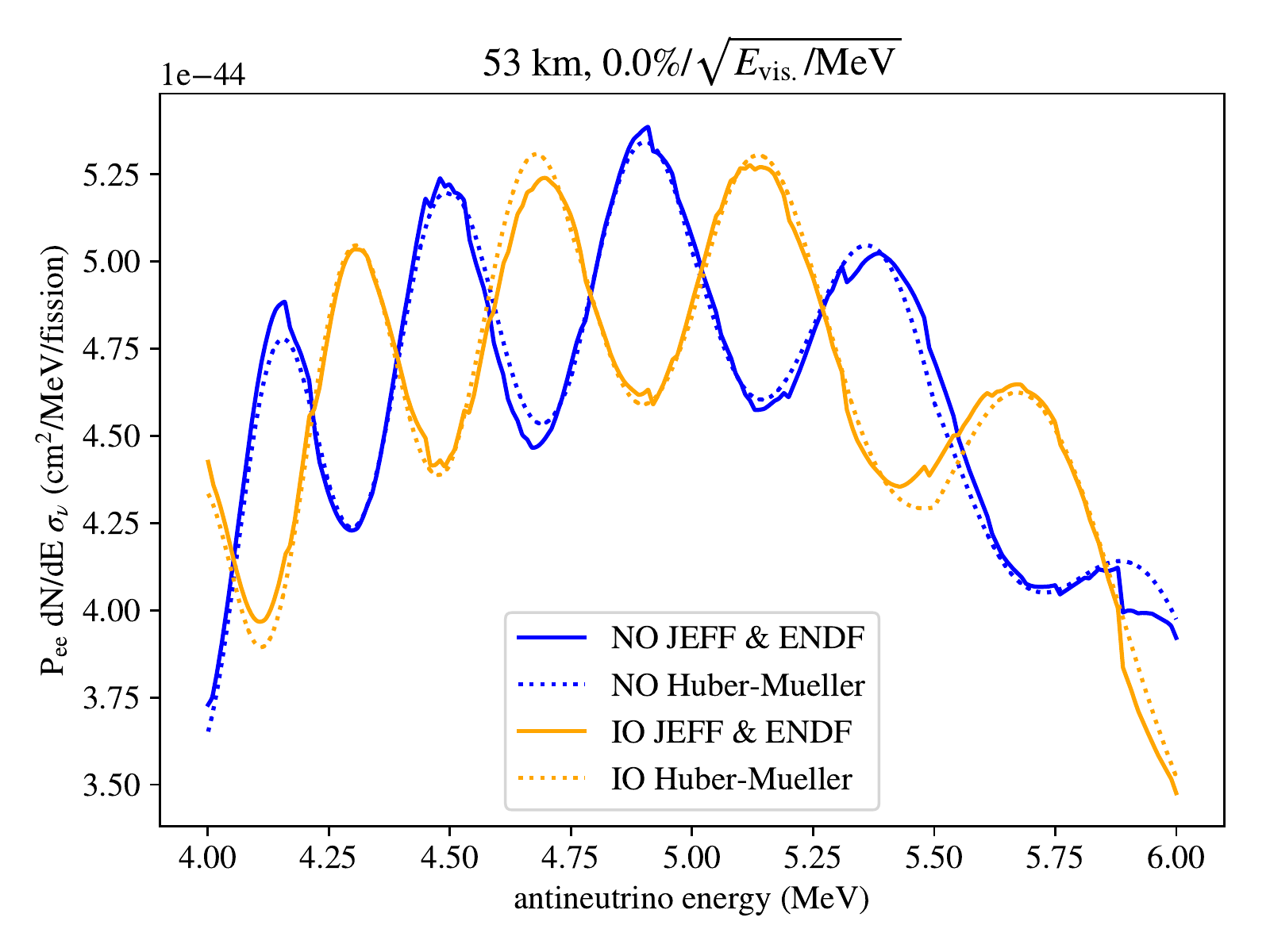}

\caption{The JEFF-3.1.1 \& ENDF/B-VIII.0 antineutrino spectrum per fission, and magnified sections, at 53~km from the reactor for normal order (NO) and inverted order (IO) mass hierarchies, overlaid with the respective Huber-Mueller spectra for comparison. Perfect energy resolution is assumed.
The oscillation parameters were taken to be the best fit values obtained by de Salas {\it et al.} \cite{deSalas}:
$\delta m^2\equiv\Delta m_{21}^2=7.55\times 10^{-5} \mathrm{eV}^2$, 
$\Delta M^2\equiv\frac{1}{2}[\Delta m_{31}^2+\Delta m_{32}^2]=+2.46_{-0.03}^{+0.03}\:(-2.46_{-0.03}^{+0.04})\times 10^{-3} \mathrm{eV}^2$,
$\sin^2\theta_{12}=0.320$, and $\sin^2\theta_{13}=0.02160\:(0.02220)$ for the NO (IO) hierarchy.
The magnified views exemplify the sawtooth distortions that run over almost the entire spectrum. Yet in any given energy window, the magnitude of the sawtooth distortion is small relative to the mass hierarchy-dependent oscillations.}\label{fig:spectra}
\end{figure}
\begin{figure}
\includegraphics[width= \linewidth]{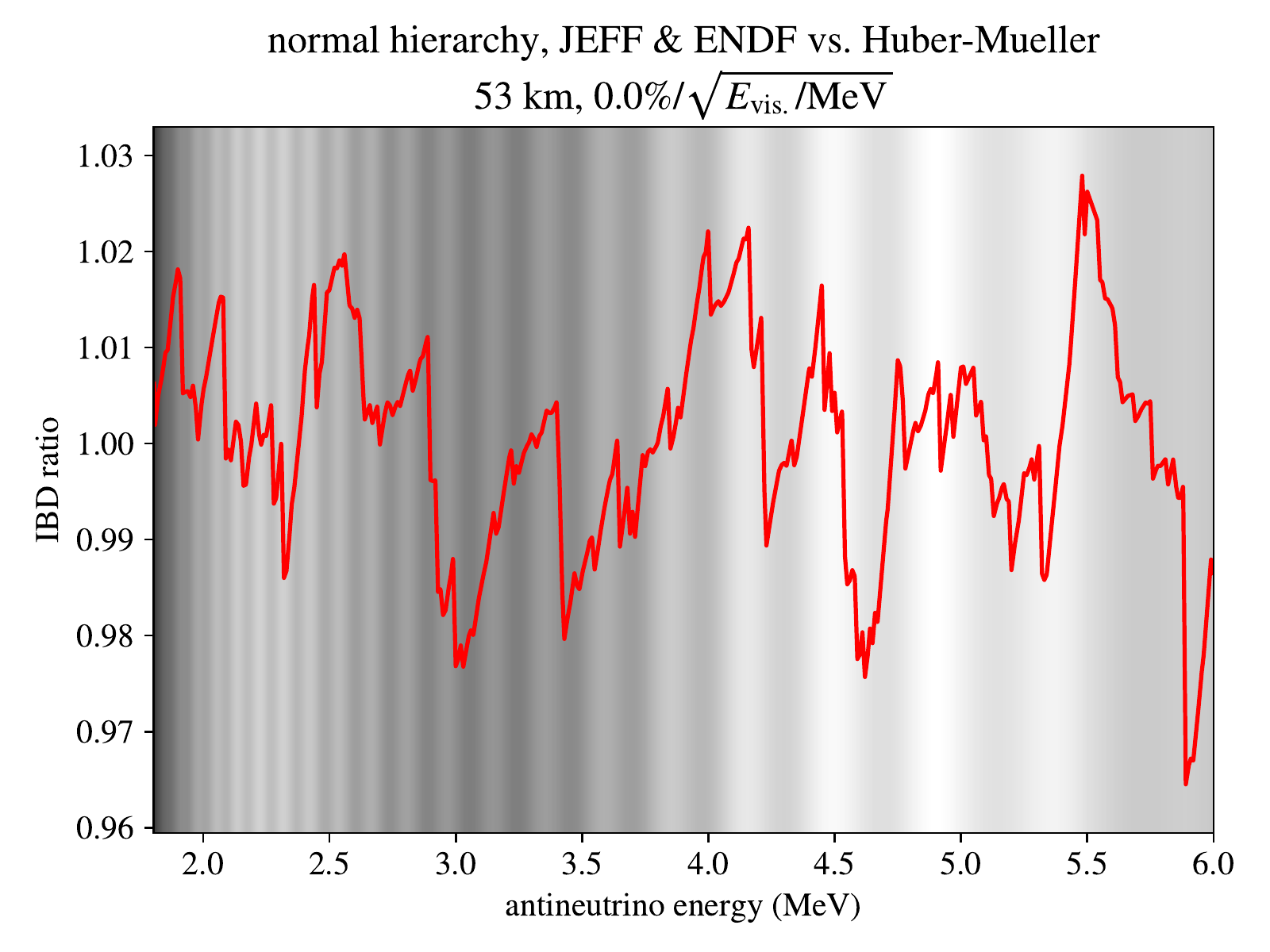}
\includegraphics[width= \linewidth]{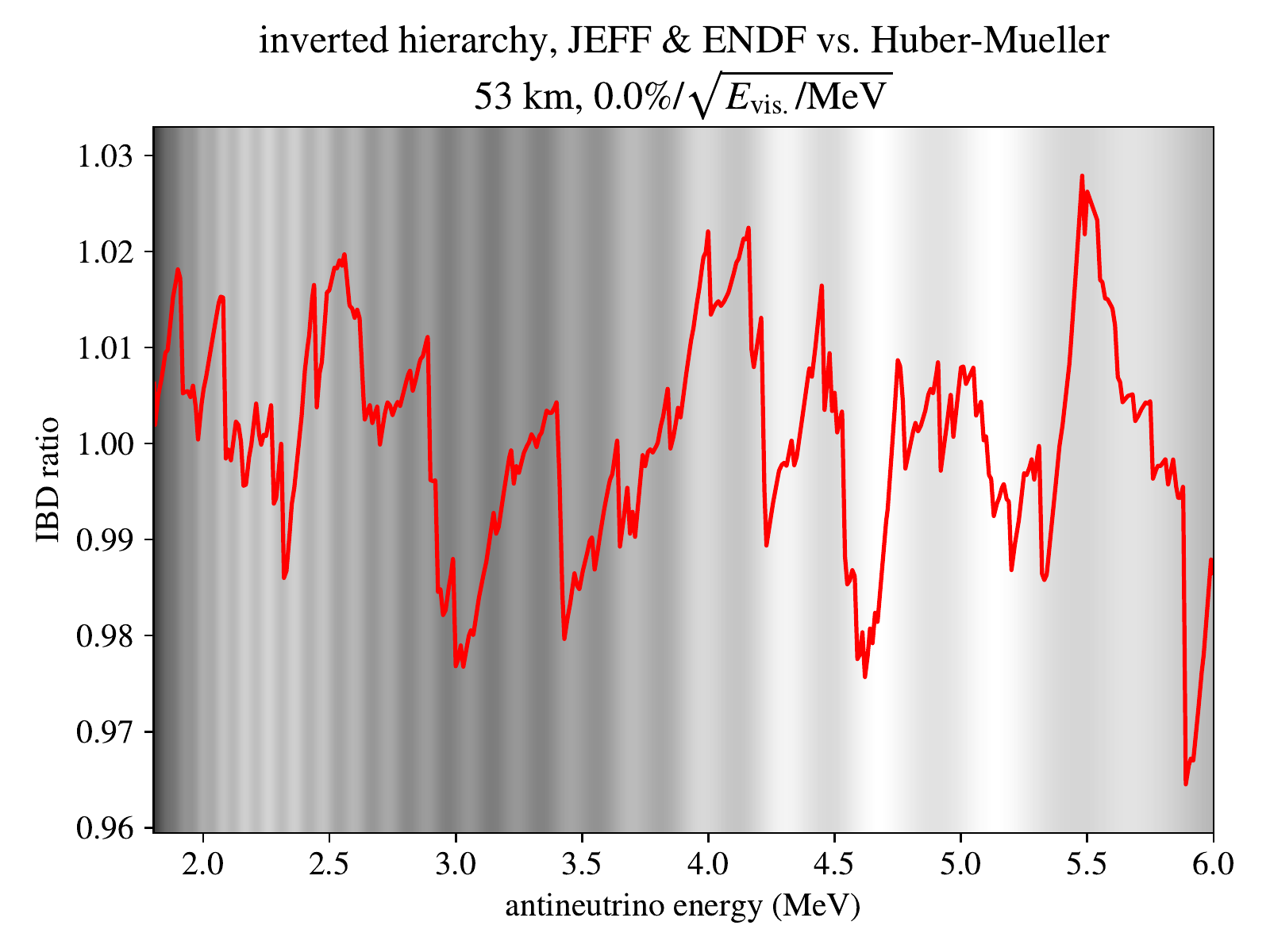}
\caption{The ratio of the JEFF~\&~ENDF spectrum to the Huber-Mueller spectrum for antineutrinos per fission at 53~km from the reactor, for the normal ordered hierarchy on the top and inverted on the bottom. This plot truncates at 6~MeV to avoid downscaling the abscissa for broad shape differences between the spectra, which dominate over the sawtooth distortions above this energy as is visible in Figure~\ref{fig:spectra}. The oscillation parameters were taken to be their central values as given in Figure~\ref{fig:spectra}. By dividing out the flux scale, this view exaggerates the apparent severity of the sawtooth distortions throughout the spectrum, including some that may align with $\Delta m^2_{3i}$ oscillations. Yet in any given energy window, the magnitude of the sawtooth distortion remains small, as is better seen in Figure~\ref{fig:spectra}.}
\label{fig:ratio}
\end{figure}

\section{Primary Analysis}\label{sec:primary}
\subsection{Beta Decay Structure of the\\Reactor Antineutrino Spectrum}\label{sec:beta}

Small sawtooth-like structures are predicted by summation calculations \cite{Hayes-Bump}, 
in which the 
aggregate fission antineutrino spectrum is constructed from the sum over
all individual beta decays of the fission fragments 
weighted by the corresponding cumulative fission yields.
The origin of sawtooth structures in fission antineutrino spectra is discussed in detail in reference~\cite{fine-structure}. 
For an individual beta decay transition, the Fermi function, needed to account for
 the interaction of the outgoing electron with the charge of the 
daughter nucleus, results in an abrupt cutoff in the antineutrino spectrum at the endpoint energy. As a result aggregate fission antineutrino spectra, which represent a sum of individual spectra from the decays of many fission fragments, 
exhibit sawtooth structures which become pronounced when the density of endpoints in a given energy window is low.
Sonzogni {\it et al.} \cite{fine-structure} have used this fact to reveal dominant contributions of particular fission fragments 
to the 
 measured 
antineutrino spectrum from the Daya Bay experiment \cite{DayaBay}.
The fission antineutrino spectra often used as the `expected' spectra in oscillation analyses, such as the Huber-Mueller \cite{huber, mueller} spectra, 
do not exhibit any sawtooth structure because they have been smoothed by averaging over broad energy bins, typically of width 
$\sim$250~keV; the situation is summarized in reference~\cite{fine-structure}.

Recently Forero {\it et al.} \cite{forero} have raised a concern that the actual sawtooth structures, along with the uncertainties in their magnitude and energy 
positions, present a serious challenge to the JUNO mass hierarchy experiment---one that suggests the need for a near detector
of similar energy resolution to
the main detector at 53~km. To address this concern, and since the main JUNO baseline is nearly optimal for this type of experiment, we adopt a 53~km reactor-detector distance in our analysis. We will demonstrate that the sawtooth structures are negligibly small, and that even if nature chooses oscillation parameters resulting in a near degeneracy between the mass hierarchies' oscillation patterns, the uncertainty in the detailed structure of the reactor spectrum is insufficient to impede hierarchy measurement with an experiment like JUNO if energy response nonlinearities are controlled to the requisite degree \cite{dayabay2, qian}.

The contribution of each individual decay spectrum into the total spectrum, 
which is determined by both the 
corresponding beta decay branching ratios and the cumulative fission yield, is generally not well known.
Thus, the actual sawtooth distortions of the antineutrino spectrum may differ from those predicted by the summation calculations used.
Section~\ref{sec:decayUncertainties} addresses these uncertainties, but as starting points we use the JEFF-3.1.1 (Joint Evaluated Fission and Fusion) nuclear database \cite{JEFF} and the ENDF/B-VIII.0 (Evaluated Nuclear Data File) decay data sub-libraries \cite{ENDF} for cumulative fission yields and beta decay branching ratios, respectively. The resulting spectrum is hereafter denoted ``JEFF~\&~ENDF.''

Together these databases predict \cite{hayes-evolution} a small, but statistically nonsignificant, 
anomaly relative to the Daya Bay measurements \cite{DayaBay}, and provide a good description
of the changes in the antineutrino energy spectrum due to the evolution of the reactor fuel \cite{DayaBay-new}.
In Figure~\ref{fig:spectra} we show the JEFF~\&~ENDF prediction of the inverse-beta-decay (IBD) spectrum expected given perfect energy resolution at 53~km from a reactor with effective isotopic fission fractions of 57.1\%~\isotope[235]{U}, 7.6\%~\isotope[238]{U}, 29.9\%~\isotope[239]{Pu}, and 5.4\%~\isotope[241]{Pu}, comprising the time averaged fractions reported by Daya Bay \cite{DayaBay-new}.

To isolate the effect of the sawtooth features, we use the Huber-Mueller \cite{huber, mueller} spectrum as a standard of comparison with the same effective fission fractions. Since the Huber-Mueller spectrum covers only antineutrino energies from 2~MeV through 8~MeV, we continue the spectrum outside this region using a smoothed and normalized version of the JEFF~\&~ENDF spectrum, with the smoothing applied across 250~keV bins for consistency with the Huber-Mueller spectrum itself. We also rescale every detailed spectrum compared to this Huber-Mueller spectrum so they yield the same total antineutrinos per fission when integrated from 2~MeV through 8~MeV. Figure~\ref{fig:spectra} shows the detailed JEFF~\&~ENDF spectrum overlaid against the Huber-Mueller spectrum.

Figure~\ref{fig:ratio} plots the ratio of the JEFF~\&~ENDF prediction divided by the Huber-Mueller model of the reactor antineutrino spectrum, thus exaggerating the appearance of small differences between the two spectra. Sawtooth-like features are clearly visible, as is a quasi-periodic component in their structure. The same features are visible as jagged edges in the JEFF~\&~ENDF curves in Figure~\ref{fig:spectra}, and are seen to be small relative to the antineutrino disappearance oscillation amplitudes. In the background of Figure~\ref{fig:ratio}, following Forero {\it et al.} \cite{forero}, we shade the contour of the Huber-Mueller spectrum alone, illustrating possible alignments between the sawtooth features and the mass hierarchy-dependent oscillations themselves. This view raises the question of whether the details of the reactor-spectral structure might impede experiments aiming to determine the neutrino mass ordering from small disappearance oscillations modulating that spectrum \cite{forero}. As will be shown, however, Figure~\ref{fig:spectra} puts the fine structure of the reactor spectrum into correct perspective, as negligible relative to the hierarchy-dependent oscillation pattern.

Having simulated the sawtooth-like features in the detailed reactor spectra, the next subsection carries the analysis into the Fourier domain. This will prove useful in Section~\ref{sec:hierarchyDegeneracies}, addressing cases wherein the mass hierarchies appear nearly degenerate in the energy domain.

\subsection{Fourier Analysis of the\\*Detectable $L/E$ Spectrum}\label{sec:fourier}

As pointed out by Learned {\it et al.} \cite{learned} and explored by subsequent authors \cite{zhan}, a natural analysis of mass hierarchy-dependent oscillations arises in the Fourier transform of the antineutrino spectrum, when that spectrum is expressed as a function of the reactor-detector flight distance divided by antineutrino energy, or $L/E$.

With even approximate prior knowledge of the mass hierarchy-dependent oscillation frequencies, potentially confounding features can be filtered out. Although the complete Fourier transform conveys no additional information beyond what is already contained in the untransformed signal, the transform does ease the development of an analysis resilient to extraneous features such as the fine structure in the reactor spectrum, when focusing on a specific window in frequency space as following sections will demonstrate.

We present our findings in $|\Delta m^2|$ frequency space to facilitate direct comparison with existing and ongoing preparations for mass hierarchy experiments using reactor antineutrinos. This space relates to the standard inverse-$L/E$ Fourier space by,
\begin{equation}
|\Delta m^2| \equiv \frac{1}{2\alpha} \omega
\end{equation}
where $\omega$ is angular frequency with dimensions of $[L/E]^{-1}$, and $\alpha$ is the constant coefficient found in all arguments of the form $\alpha \Delta m^2_{ij} L/E$ in the neutrino flavor survival probability. The factor of $1/2$ comes from the fact that $|\Delta m^2|$ is an apparent frequency of sine-squared rather than pure sinusoid oscillations.

As will be shown, a frequency-domain analysis may prove requisite for mass hierarchy determination in cases of nearly hierarchy-degenerate combinations of oscillation parameters. For this purpose, Fourier cosine (FCT) and sine (FST) transforms are used:

\begin{equation}
F(\omega) \equiv 2 \Delta \frac{L}{E} \sum_i g\left(\omega \left[ \frac{L}{E}\right]_i \right) f\left(\left[ \frac{L}{E}\right]_i\right)
\end{equation}
where $F$ is the Fourier cosine or sine transform of $f$, $g$ denotes $\sin$ or $\cos$ according to the transform, $\Delta\frac{L}{E}$ is the $L/E$ bin width, and $\left[\frac{L}{E}\right]_i$ is the $i^\mathrm{th}$ $L/E$ bin center.

\begin{figure}
\includegraphics[width= \linewidth]{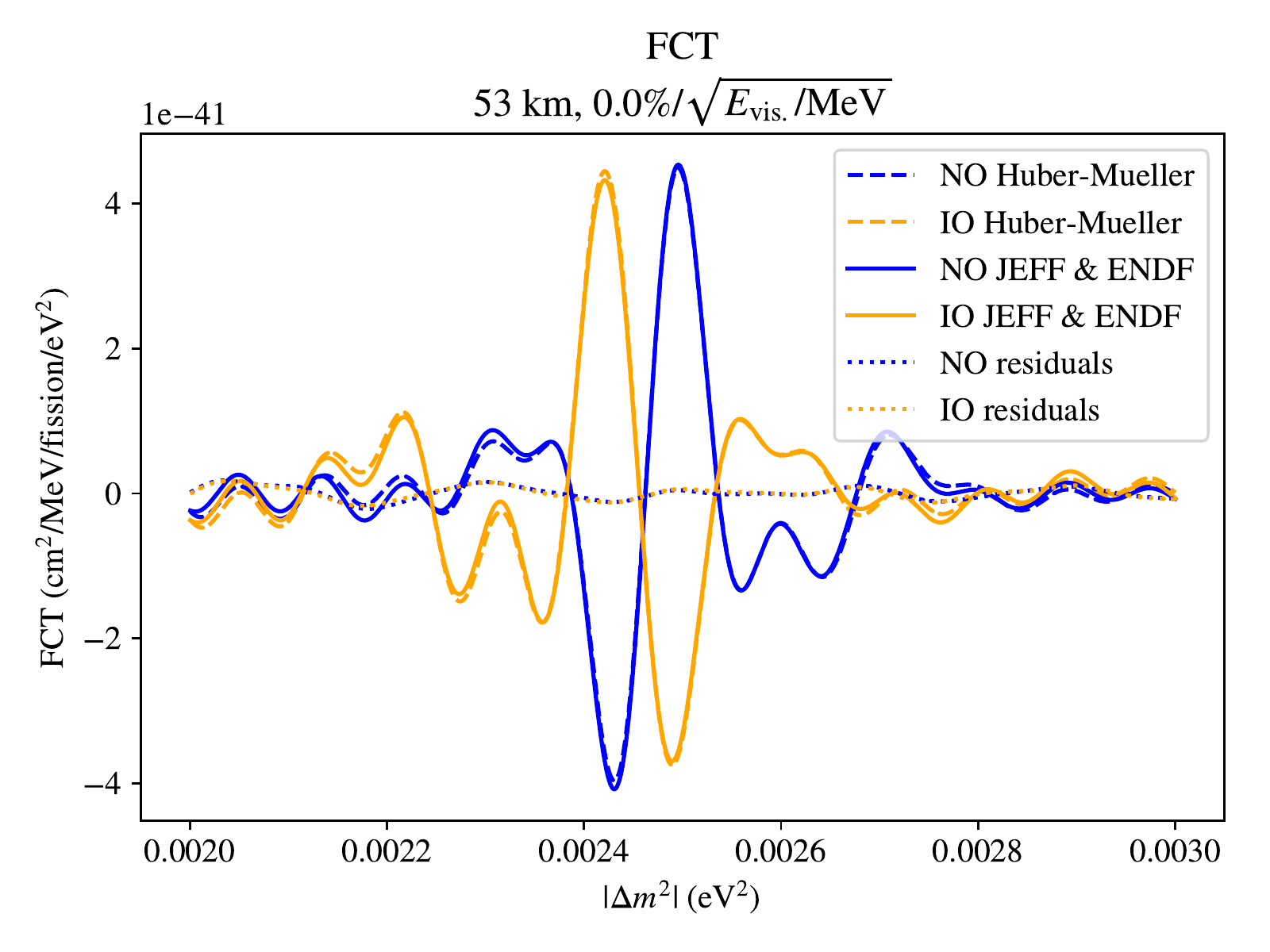}  % FCT
\includegraphics[width= \linewidth]{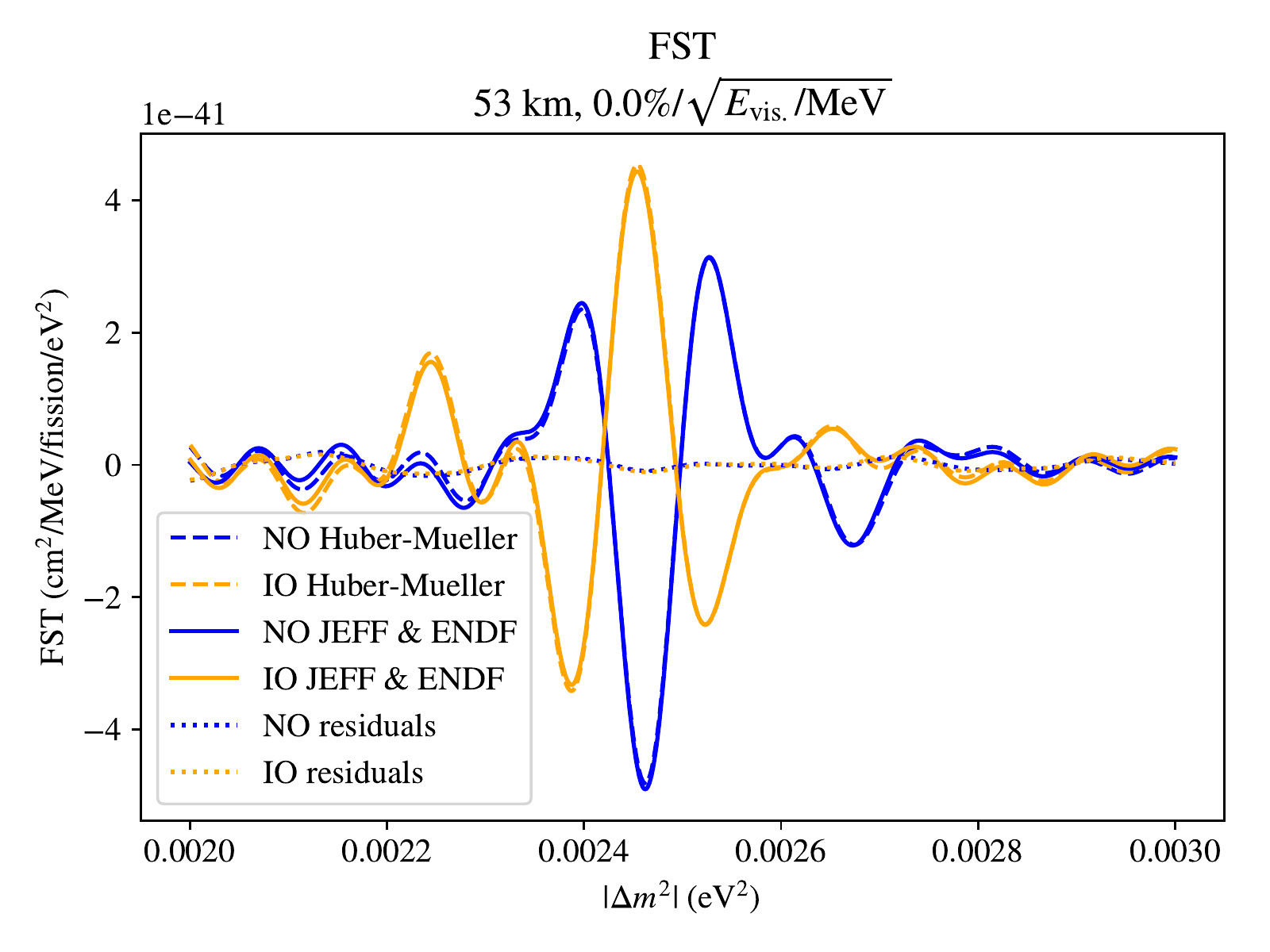}  % FST
\caption{The Fourier cosine (FCT) and sine (FST) transforms of the JEFF~\&~ENDF and Huber-Mueller antineutrino spectra per fission given perfect energy resolution at 53~km from the reactor, for normal order (NO) and inverted order (IO) mass hierarchies. Both are plotted as a function of the apparent oscillation frequency $|\Delta m^2|$ as defined in (1). Also shown are the residuals between the JEFF~\&~ENDF and Huber-Mueller transforms. The oscillation parameters were taken to be their central values as given in Figure~\ref{fig:spectra}. 
The nonzero residuals identify the effect of the sawtooth features, and also show it to distort the Fourier spectra by no more than a few percent of the $\Delta m^2_{31}$ peak amplitude in the mass hierarchy-sensitive region.
}
\label{fig:fourier}
\end{figure}

Figure~\ref{fig:fourier} shows the Fourier sine and cosine transforms of the JEFF~\&~ENDF and Huber-Mueller spectra, as a function of the apparent $|\Delta m^2|$ frequency in the region around $\Delta m^2_{31}$. Also shown are the residuals obtained by subtracting the transformed Huber-Mueller spectrum from the transformed JEFF~\&~ENDF spectrum. These residuals indicate that the sawtooth-like features distort the Fourier spectra by no more than three percent of the $\Delta m^2_{31}$ peak amplitude, in the region of oscillation frequency space relevant to distinguishing a mass hierarchy.

\begin{figure}
\includegraphics[width=0.7\linewidth]{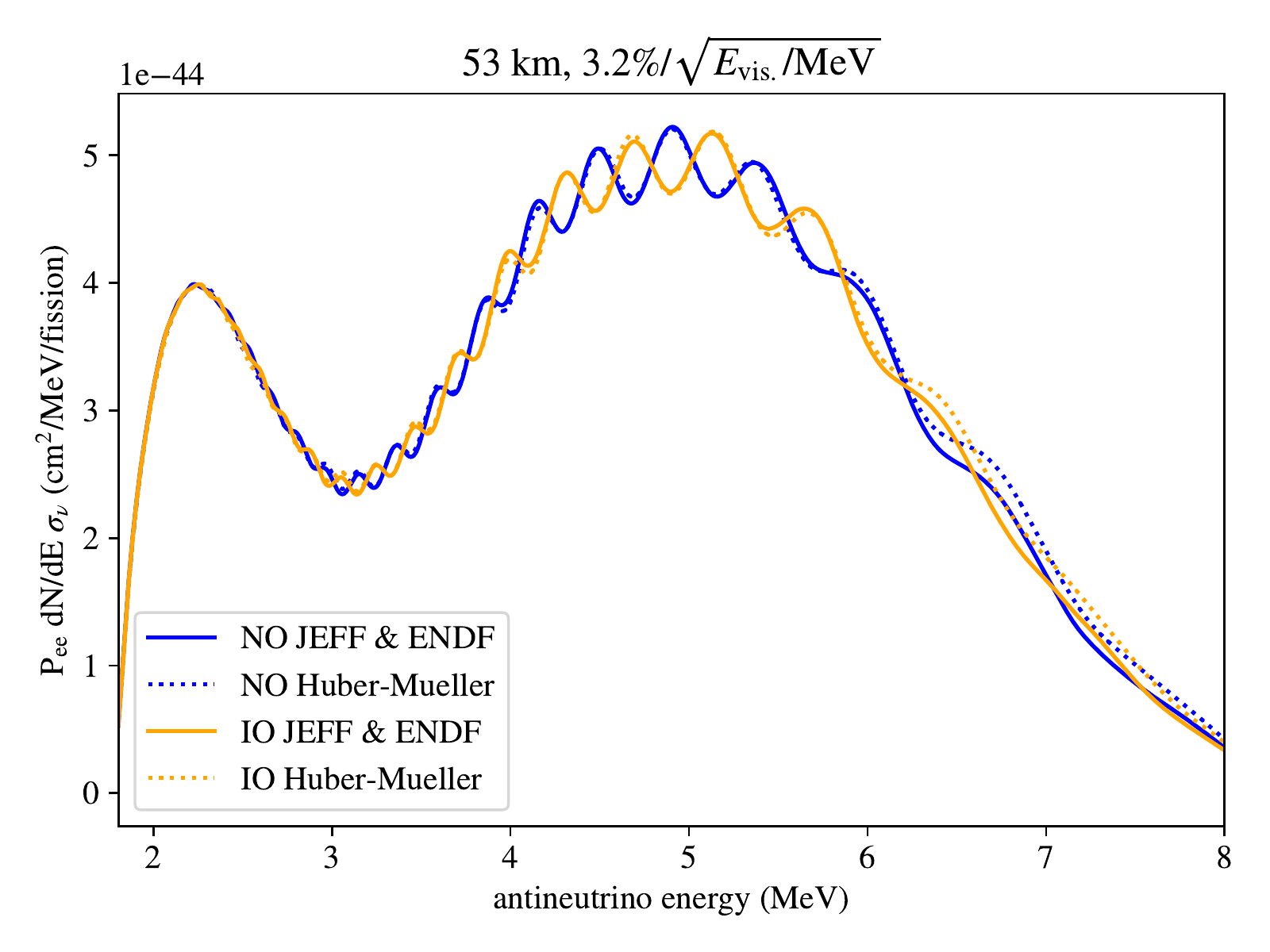} % Smeared E
\includegraphics[width=0.7\linewidth]{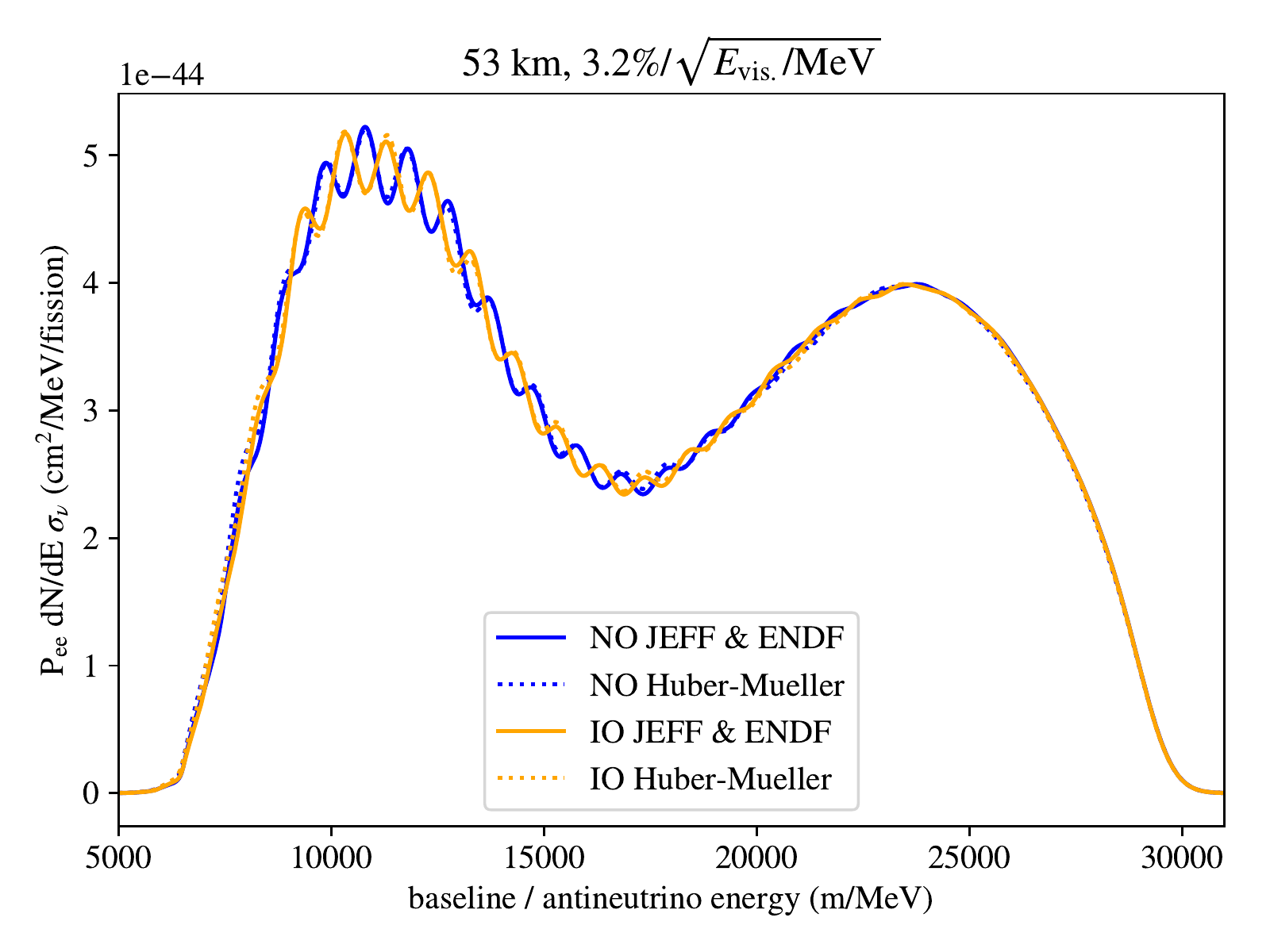} % Smeared L/E
\includegraphics[width=0.7\linewidth]{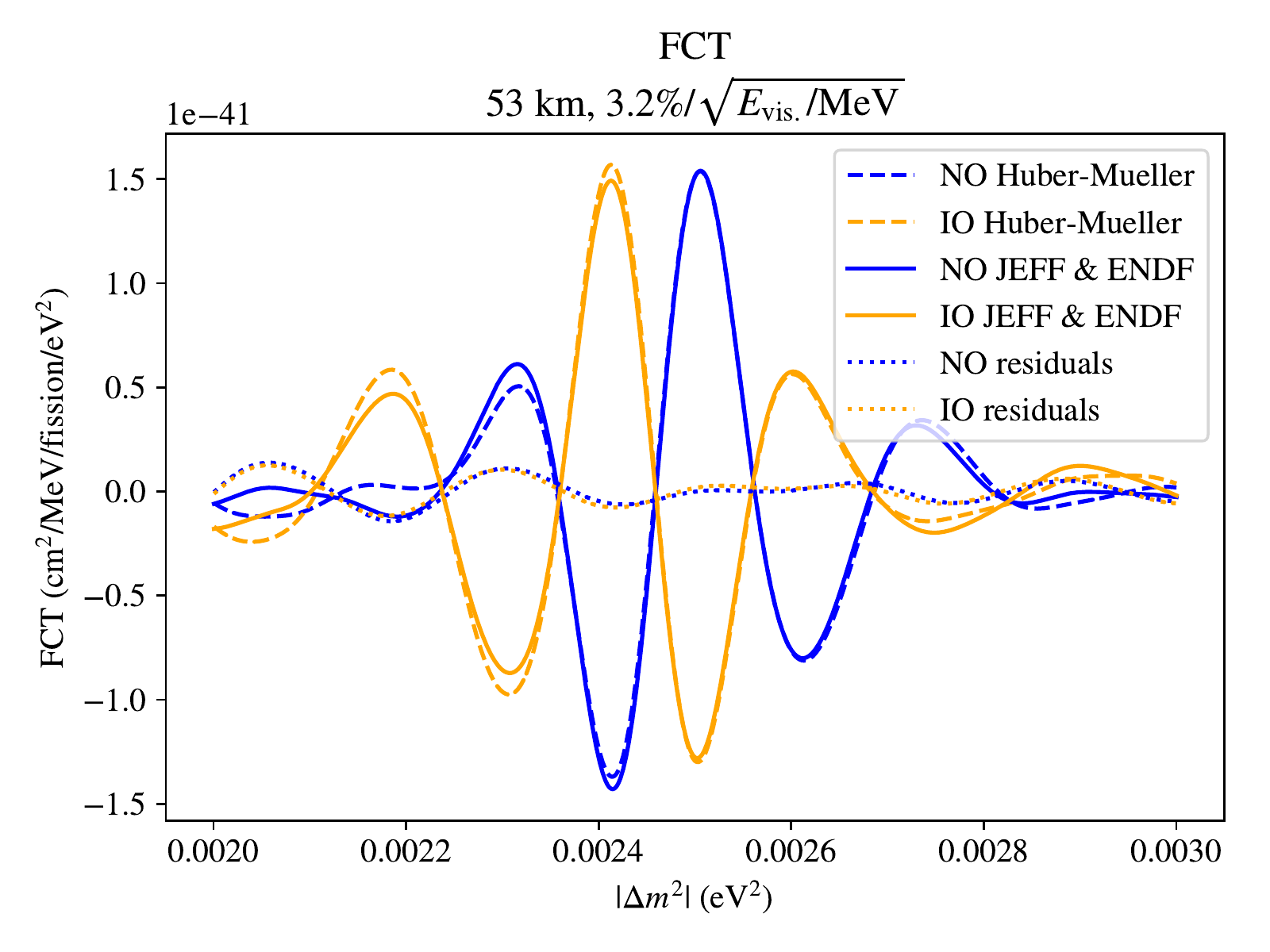} % Smeared FCT
\includegraphics[width=0.7\linewidth]{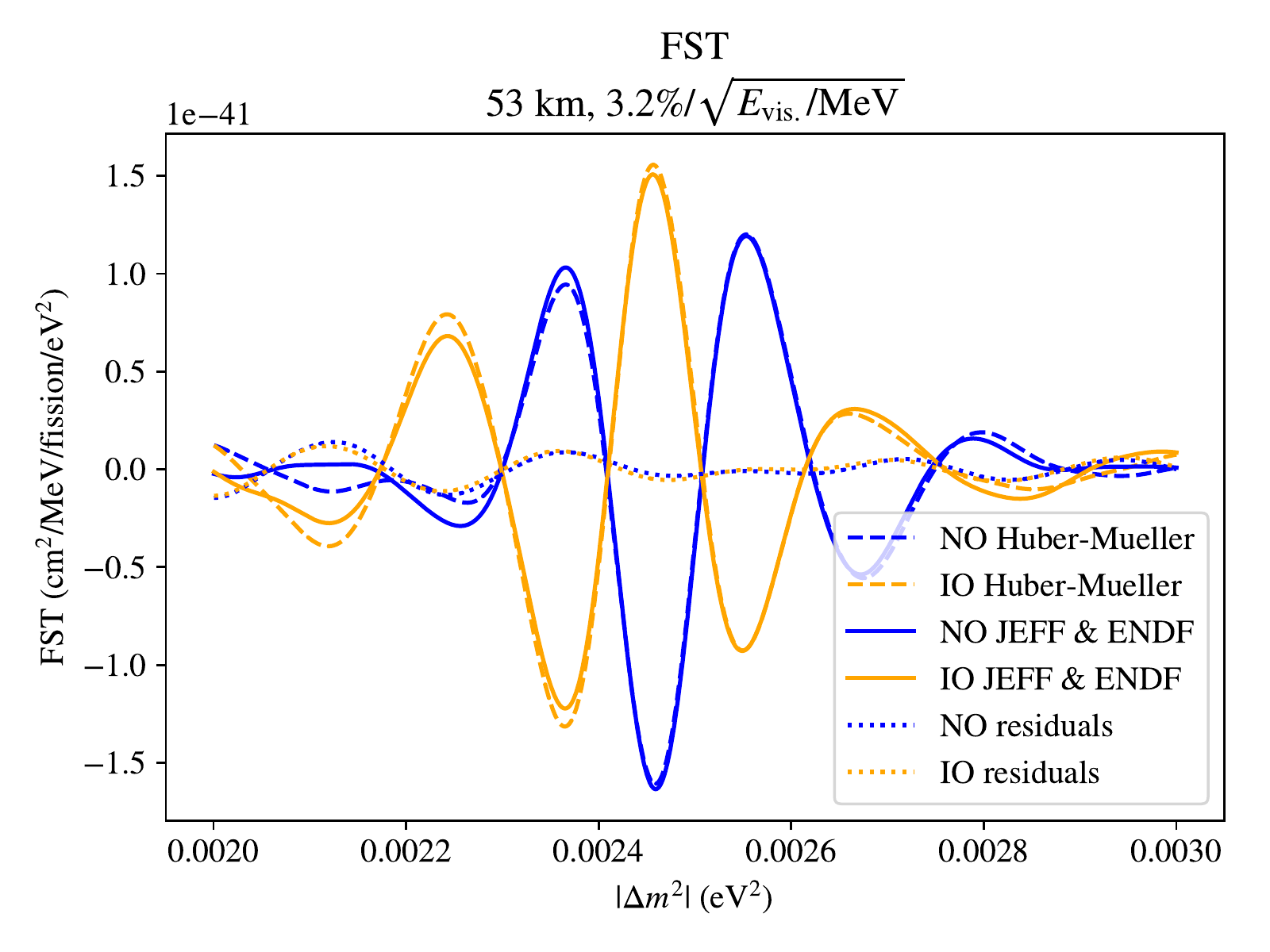} % Smeared FST
\caption{
The JEFF~\&~ENDF and Huber-Mueller antineutrino spectra per fission are shown along the top, for normal order (NO) and inverted order (IO) mass hierarchies given $3.2\%/\sqrt{E_\mathrm{vis.}/\mathrm{MeV}}$ energy resolution at 53~km from the reactor. The first graph shows these spectra as a function of energy while
the second graph shows them as a function of flight distance divided by energy ($L/E$).
Below them are their Fourier cosine (FCT) and sine (FST) transforms and the residuals between the JEFF~\&~ENDF and Huber-Mueller transforms. The oscillation parameters were taken to be their central values as given in Figure~\ref{fig:spectra}.
	As was the case for perfect energy resolution, the nonzero residuals identify the effect of the sawtooth features, but also show it to distort the Fourier spectra by only a few percent of the $\Delta m^2_{31}$ amplitude in the mass hierarchy-sensitive region.
\label{fig:smeared}
}
\end{figure}

As a consequence of imperfect detector energy resolution, the theoretical expectation for the antineutrino spectrum smears in both energy and frequency space. Figure~\ref{fig:smeared} shows the effect of $\delta E_\mathrm{vis.}/E_\mathrm{vis.}=3.2\%/\sqrt{E_\mathrm{vis.}/\mathrm{MeV}}$ energy resolution on the predicted antineutrino energy spectrum and the resulting Fourier transforms. As before, the sawtooth features in the underlying reactor spectrum distort the detectable spectrum by no more than three percent across the whole energy domain. The Fourier transform distortions remain negligible in the region sensitive to a mass hierarchy, comprising perturbations of no more than six and two tenths of one percent relative to the $\Delta m^2_{31}$ peak amplitude. As discussed in Section~\ref{sec:osc}, this effect remains limited in a similar scale throughout the domain of apparent oscillation frequencies above $10^{-3}\mathrm{eV}^2$, and we have found this result consistent across a range of detector energy resolutions from $3\%/$ through $6\%/\sqrt{E_\mathrm{vis.}/\mathrm{MeV}}$.

The next section discusses the effect of varying the dominant sources of uncertainty in our prediction, and investigates whether certain natural values of these parameters, constrained around their respective error margins, would produce circumstances where the sawtooth-like spectral structures could become an impediment to mass hierarchy determination.

\section{Uncertainties}\label{sec:uncertainties}
\subsection{Uncertainties in the\\*Reactor Antineutrino Spectrum} \label{sec:decayUncertainties}

\begin{figure}
\includegraphics[width= \linewidth]{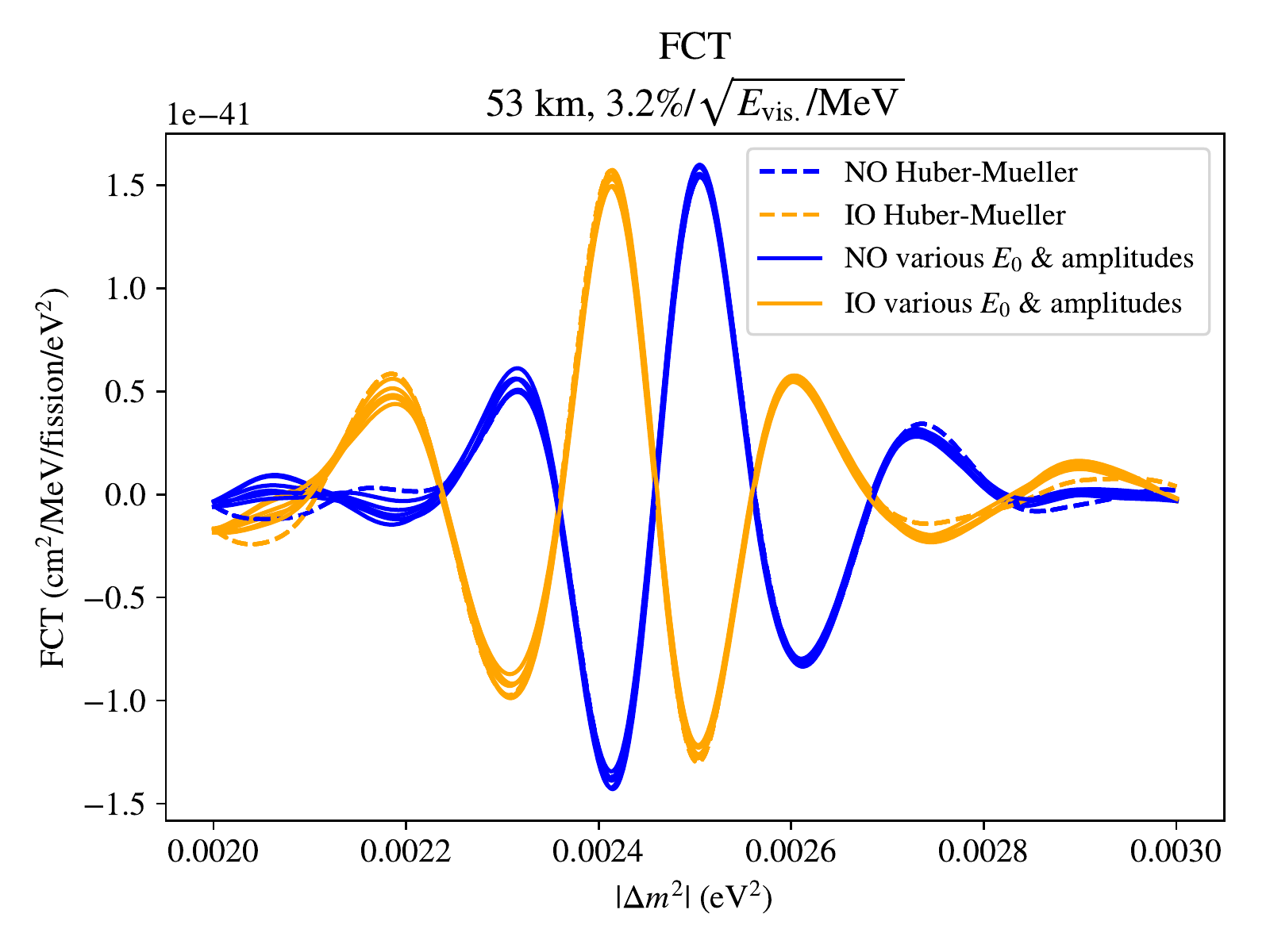}\\ % Random spectrum FCT
\includegraphics[width= \linewidth]{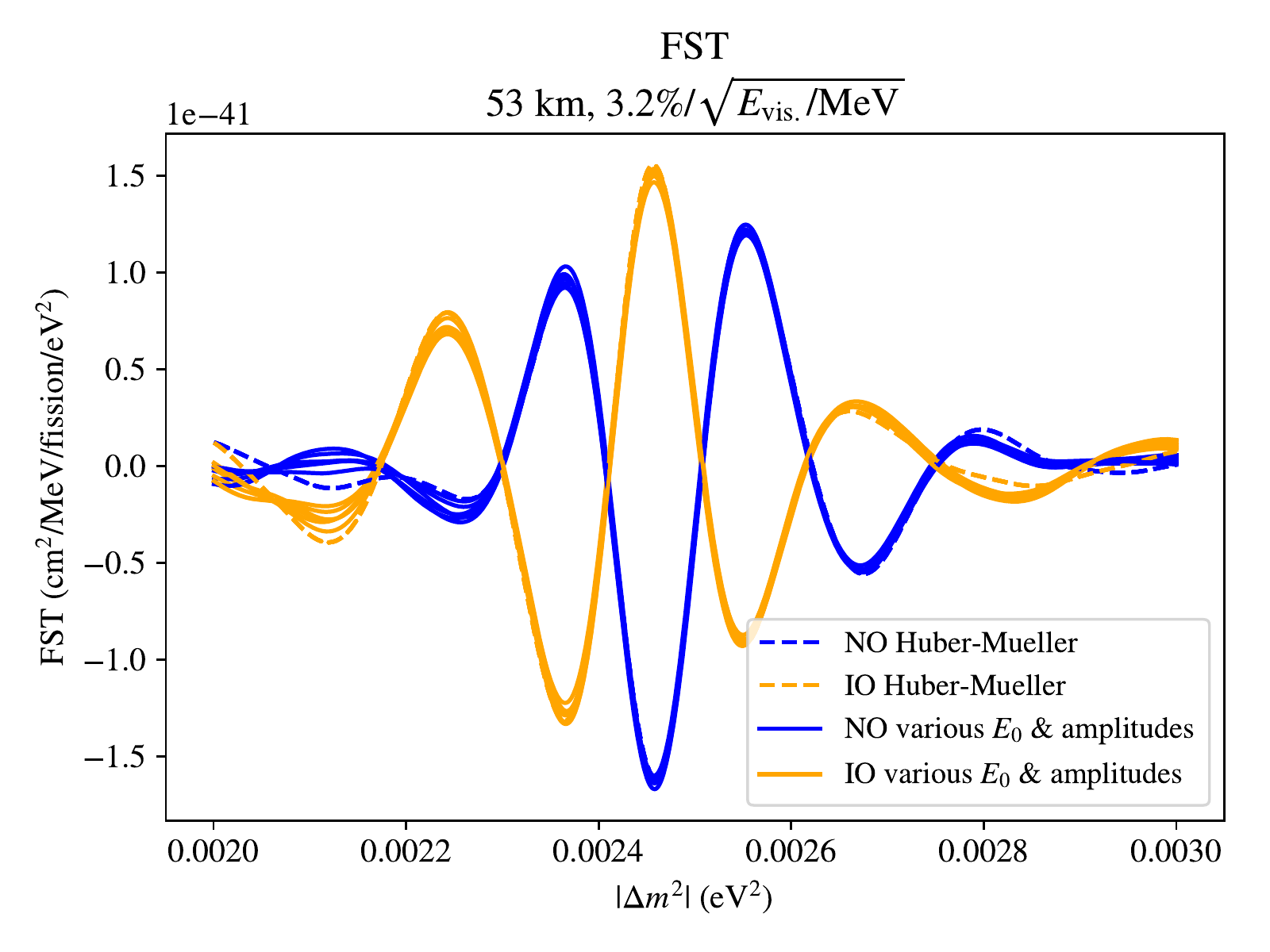} % Random spectrum FST
\caption{
The result in Fourier space of sampling random branching ratios (amplitudes) and endpoint energies ($E_0$) to generate a range of possible sawtooth patterns in the reactor spectrum, while not exceeding the maximum decay energy $Q$-value for each fission fragment. Fourier cosine (FCT) and sine (FST) transforms are shown for the resulting $L/E$ spectra and for the Huber-Mueller model for comparison, given $3.2\%/\sqrt{E_\mathrm{vis.}/\mathrm{MeV}}$ energy resolution at 53~km from the reactor. The oscillation parameters were taken to be their central values as given in Figure~\ref{fig:spectra}.
For each possible configuration of beta decays sampled, the spectra transform into a specific narrow band of curves in the mass hierarchy-sensitive region, where they distort the Fourier transforms by no more than a few percent of the $\Delta m^2_{31}$ peak. 
}
\label{fig:varied}
\end{figure}

As pointed out by Forero {\it et al.} \cite{forero} one method of examining the effect of the uncertainties in the nuclear database is to introduce 
random variations in the amplitudes of the decays, while keeping
the total IBD rate fixed.
The $Q$-values for the fission fragment beta decays, corresponding to the maximum decay energy for each fragment, are reasonably well known.
However, we assumed the branches to decays to excited nuclear states introduce random variations in the endpoint energies, while constraining decays never to exceed their $Q$-values.

This level structure of nuclear excited states is responsible for the fine structure in the reactor antineutrino spectrum, including the sawtooth-like features. In particular, as discussed in Section~\ref{sec:intro}, pronounced sawtooth features arise when the density of endpoint energies is sparse within a given energy window. By sampling a wide range of endpoint energy configurations, an array of various possible sawtooth patterns were produced, and the detectable results compared to the results of the Huber-Mueller spectrum.

Additional energy and momentum dependences modify the beta decay spectra of first forbidden transitions. These effects can be modeled as shape factors applied to the corresponding spectral components, but their impact is found to be insignificant in regard to the conclusions developed here and is therefore neglected. Further corrections to the reactor spectrum exist, but are known to be subdominant to shape factors \cite{hayes-vogel} and so are likewise neglected.

Figure~\ref{fig:varied} shows the result of varying endpoint energies and respective amplitudes underlying the reactor neutrino spectrum, i.e., varying the specific sawtooth patterns present. In every case, the sawtooth features' contribution to the Fourier spectra remains limited to the level of a few percent of the $\Delta m^2_{31}$ amplitude throughout the hierarchy-sensitive region. 

For the sawtooth-like features to significantly distort the hierarchy-sensitive region in Fourier space would require the spacing of beta decay endpoint energies to precisely align with the spacing of hierarchy-dependent oscillation peaks throughout the resolvable energy domain. These endpoint energies are known to within a few tens of keV for all decays contributing above 0.5\% of the total spectrum. Thus the chance of such alignment can be neglected, since the endpoint energies and oscillation parameters are sufficiently constrained by experimental knowledge to rule out such a coincidence between these totally independent systems.

Future analyses may quantify the vanishing chance of such alignment, but here it suffices to note that for each mass hierarchy, every possible sawtooth variation produced transforms into a specific narrow band of curves in the pertinent frequency region, and Figure~\ref{fig:varied} shows the width of both hierarchy's bands to be very small relative to their separation. Thus whatever fine structure exists in nature will not distort the hierarchy-sensitive region in Fourier space sufficiently to impede hierarchy determination.

The only significant Fourier components of the detailed reactor spectrum are separated from the hierarchy-sensitive region by over an order of magnitude in apparent $|\Delta m^2|$ frequency, as will be shown in Section~\ref{sec:zoomedOut}. Therefore even if the sawtooth amplitudes were much larger than considered here, the only Fourier distortions of significant magnitude would remain confined to a region of $|\Delta m^2|$ space far removed from the hierarchy-dependent features.

\subsection{Uncertainties in the\\Neutrino Oscillation Parameters} \label{sec:osc}

\begin{figure}
\includegraphics[width= 0.7\linewidth]{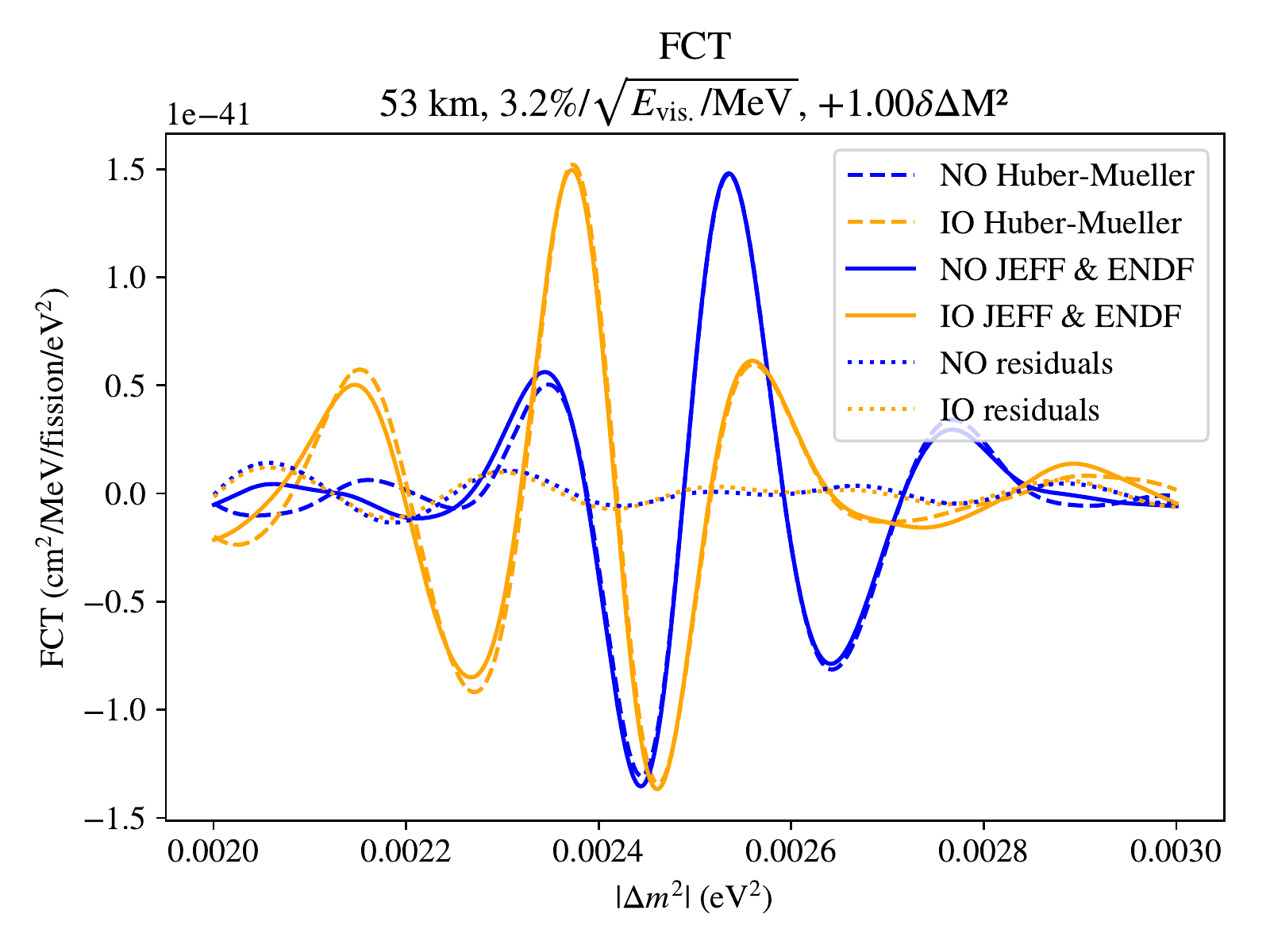} % 1sigma parameters FCT
\includegraphics[width= 0.7\linewidth]{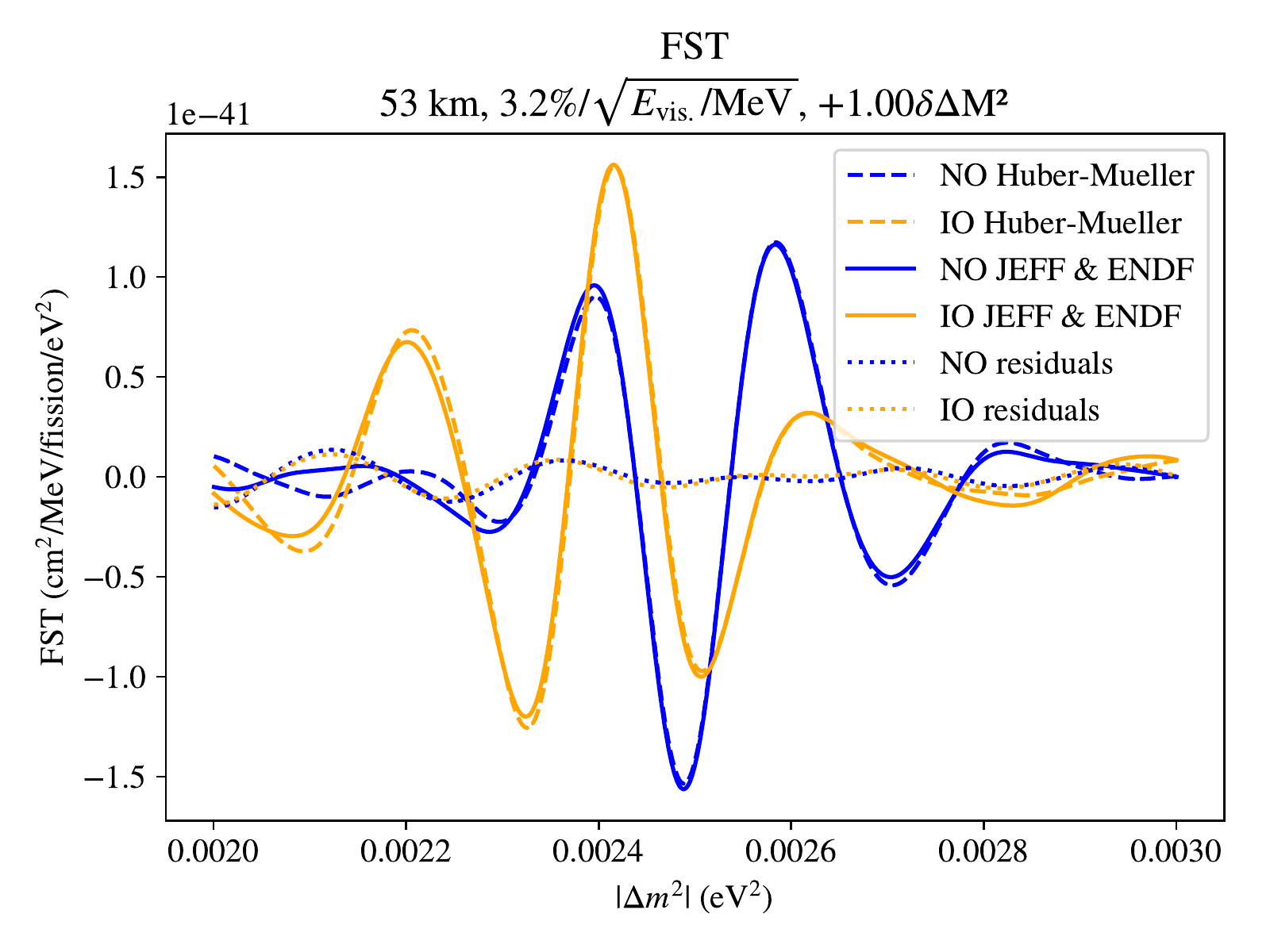} % 1sigma parameters FST
\includegraphics[width= 0.7\linewidth]{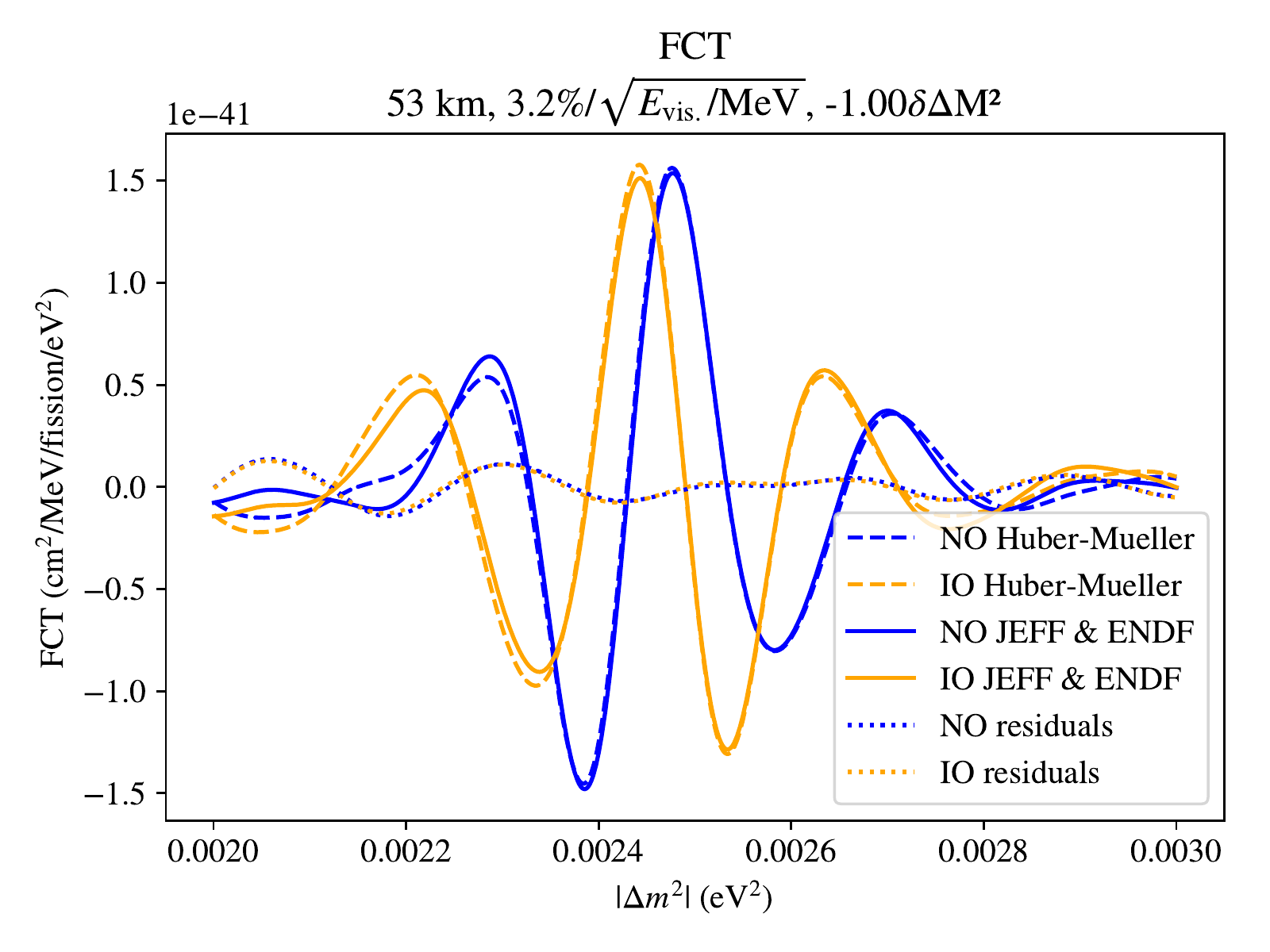} % 1sigma parameters FCT
\includegraphics[width= 0.7\linewidth]{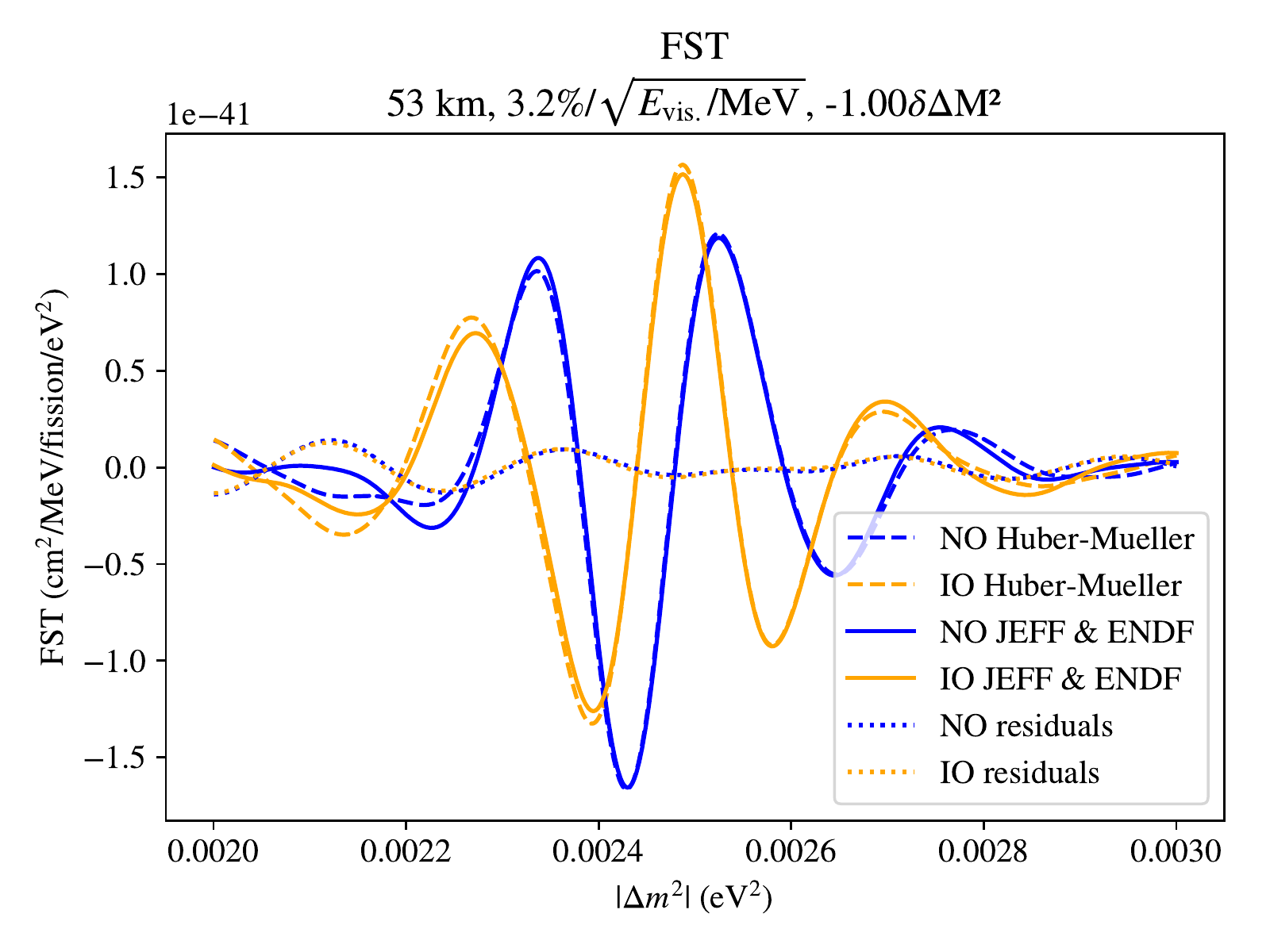} % 1sigma parameters FST
\caption{The effect in Fourier space of varying the mass hierarchy-dependent oscillation parameter $\Delta M^2\equiv\frac{1}{2}(\Delta m_{31}^2+\Delta m_{32}^2)$ to one standard deviation above and below its central values given current experimental error margins \cite{deSalas}, for normal ordered (NO) and inverted ordered (IO) mass hierarchies. Fourier cosine (FCT) and sine (FST) transforms are shown for the JEFF~\&~ENDF and Huber-Mueller antineutrino spectra per fission given $3.2\%/\sqrt{E_\mathrm{vis.}/\mathrm{MeV}}$ energy resolution at 53~km from the reactor. Also shown are the residuals between the JEFF~\&~ENDF and Huber-Mueller spectra.
	As was the case for the central values of $\Delta M^2$, the nonzero residuals identify the effect of the sawtooth features, but show it to distort the Fourier spectra by only a few percent of the $\Delta m^2_{31}$ amplitude regardless of which hierarchy parameter value is given by nature.
}
\label{fig:1sigma}
\end{figure}

\begin{figure}
\includegraphics[width= \linewidth]{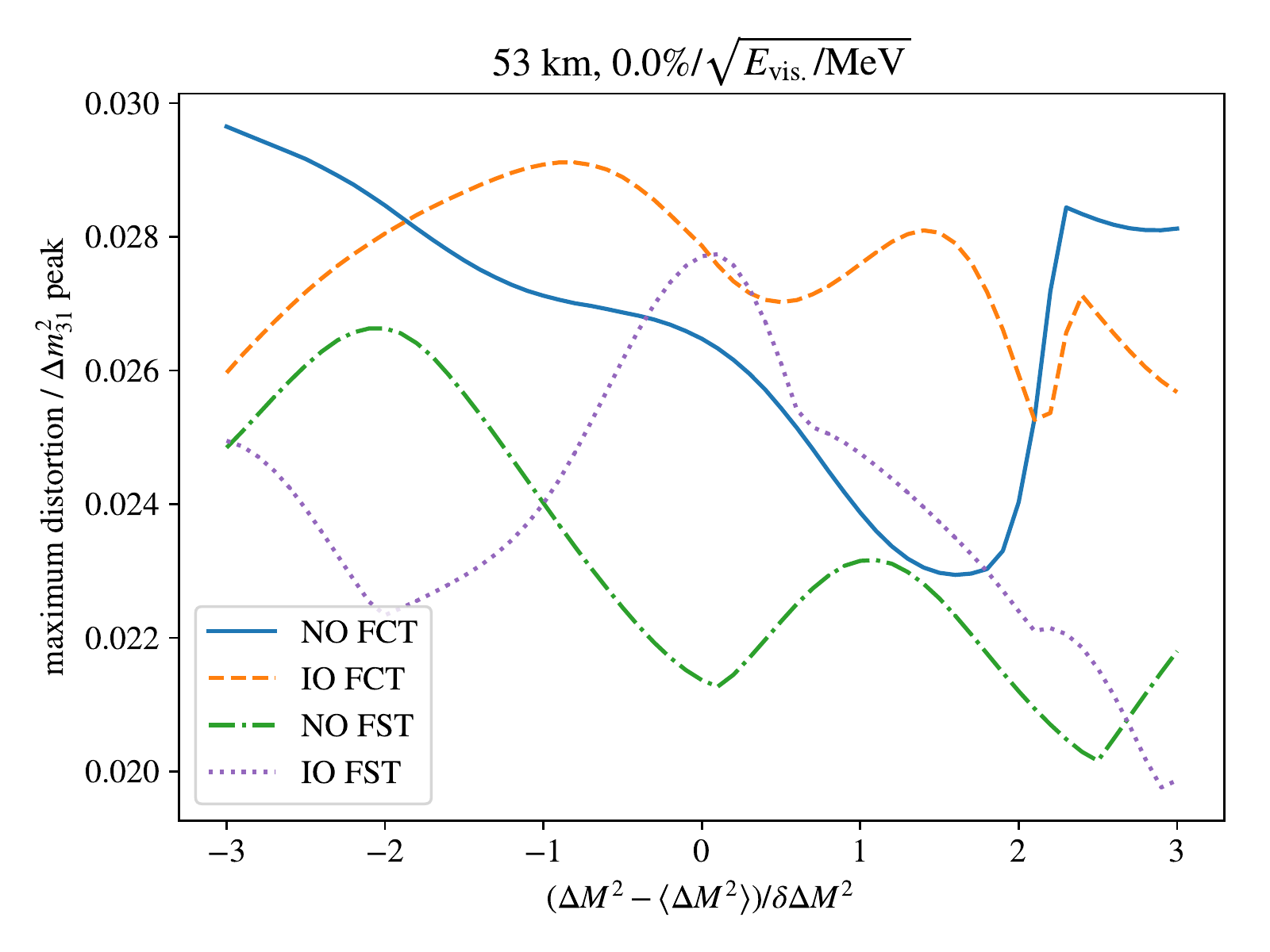} % perfect resolution scan
\includegraphics[width= \linewidth]{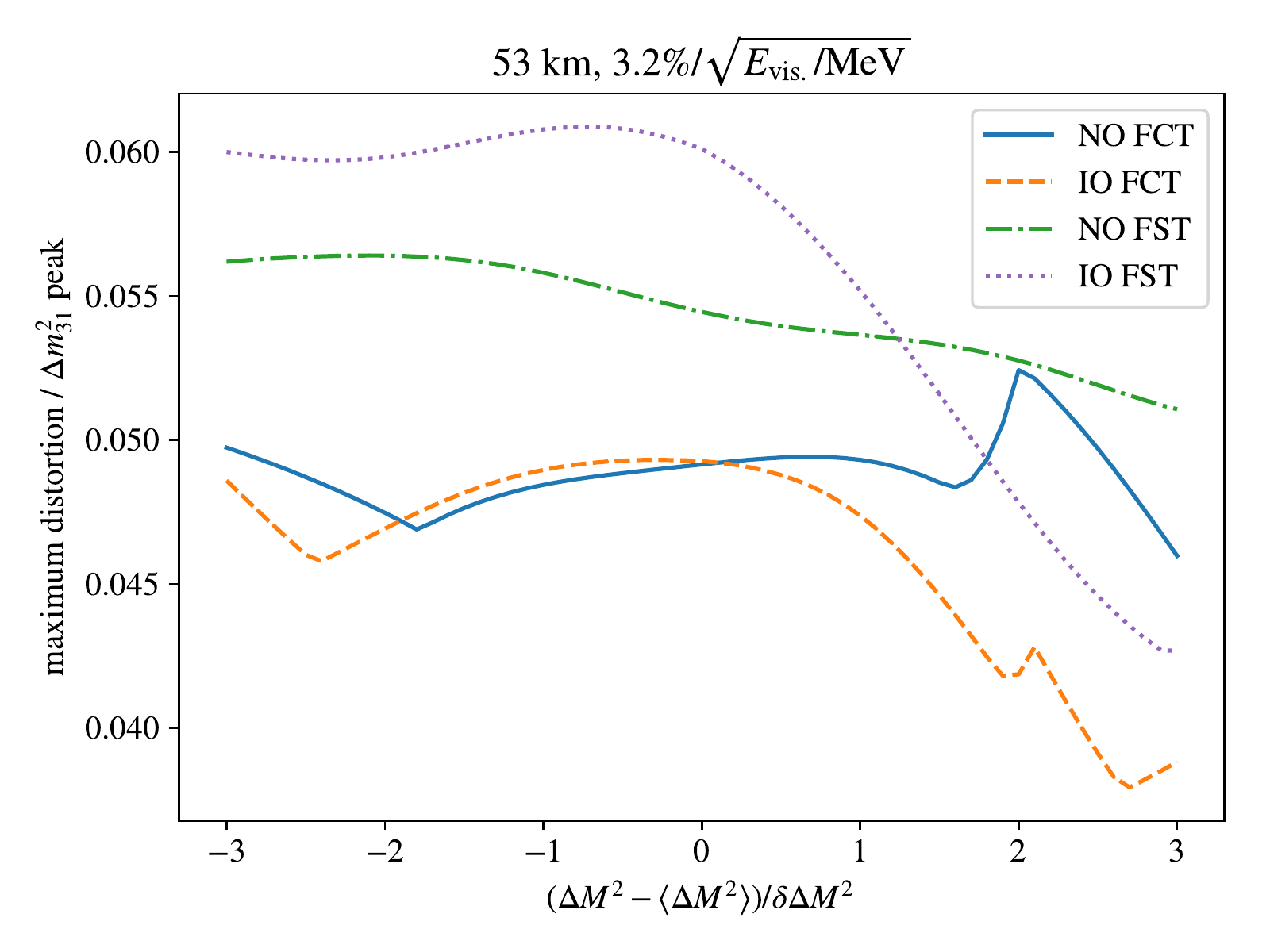} % 3.2% scan
\caption{As a fraction of the Huber-Mueller $\Delta m^2_{31}$ peak amplitude, the maximum amount by which the JEFF~\&~ENDF sawtooth features distort the Fourier cosine (FCT) and sine (FST) transforms in the hierarchy-sensitive region from 0.00234~eV$^2$ through 0.00255~eV$^2$ \cite{deSalas}, as a function of the experimentally allowed range of $\Delta M^2$ values. The ordinate indicates the deviation of $\Delta M^2$ from its central value $\langle  \Delta M^2 \rangle$, in multiples of its error margins $\delta\Delta M^2$ as given in Figure~\ref{fig:spectra}. A detector 53~km distant from a reactor is assumed, with the result of perfect energy resolution shown on the top, and $3.2\%/\sqrt{E_\mathrm{vis.}/\mathrm{MeV}}$ on the bottom. Throughout the experimentally allowed range of hierarchy-dependent oscillation frequencies, the sawtooth features distort the Fourier features critical to hierarchy determination by no more than three percent given perfect energy resolution, or six and two tenths of one percent given $3.2\%/\sqrt{E_\mathrm{vis.}/\mathrm{MeV}}$ resolution, regardless of whether the mass hierarchy is normal ordered (NO) or inverted ordered (IO).}
\label{fig:scan}
\end{figure}

Significant uncertainties remain on the magnitudes of the hierarchy-dependent mass-squared differences $\Delta m^2_{31}$ and $\Delta m^2_{32}$.
Due to the relationship among all three mass-squared differences, $ \Delta m^2_{21} = \Delta m^2_{32} - \Delta m^2_{31}$, the mass hierarchy-dependent
parameters can be consolidated into a single average value $\Delta M^2 \equiv \frac{1}{2}(\Delta m^2_{31} + \Delta m^2_{32})$ without loss of generality.
We adopt this convention, and have sampled $\Delta M^2$ values within three times its upper and lower error margins, for each respective hierarchy.
We have applied a similar procedure to the remaining oscillation parameters, and found their present uncertainties sufficiently small that we adopt their central values as constant for the purpose of this work.

Figure~\ref{fig:1sigma} shows the result of varying the hierarchy-dependent oscillation parameter $\Delta M^2$ to one standard deviation above and below its central value for each neutrino mass hierarchy. As a fraction relative to the Huber-Mueller $\Delta m^2_{31}$ peak, Figure~\ref{fig:scan} plots the maximum fraction by which the sawtooth features distort the Fourier transform in the hierarchy-sensitive region $0.00234 \le |\Delta m^2|/ \mathrm{eV}^2 \le 0.00255$, as a function of $\Delta M^2$ values ranging over three times its experimental error margin in either direction.

In every case, and for every configuration of beta decays considered, the distinct effect of the detailed spectrum is to perturb the Fourier transforms by a residual of no more than six (three) percent relative to the height of the $\Delta m^2_{31}$ peak, given $3.2\%/\sqrt{E_\mathrm{vis.}/\mathrm{MeV}}$ (perfect) energy resolution. In contrast to this effect, the features arising from the antineutrino oscillations generated by the mass-squared differences $\Delta m^2_{31}$ and $\Delta m^2_{32}$, which encode the mass ordering, dominate by over sixteen (twenty-five) times in prominence to the contribution of the sawtooth-like features from the detailed energy spectrum.

Thus the fine structure of the reactor spectrum is sure to be small relative to hierarchy-dependent oscillations. For some natural values of oscillation parameters, however, the difference between best-fit hierarchy patterns may also be small. The next section addresses the effect of the detailed reactor spectrum given nearly indistinguishable oscillation patterns between opposite mass hierarchies.

\section{Challenges}\label{sec:challenges}
\subsection{Degeneracies Between the Mass Hierarchies}\label{sec:hierarchyDegeneracies}
\begin{figure}
\includegraphics[width= \linewidth]{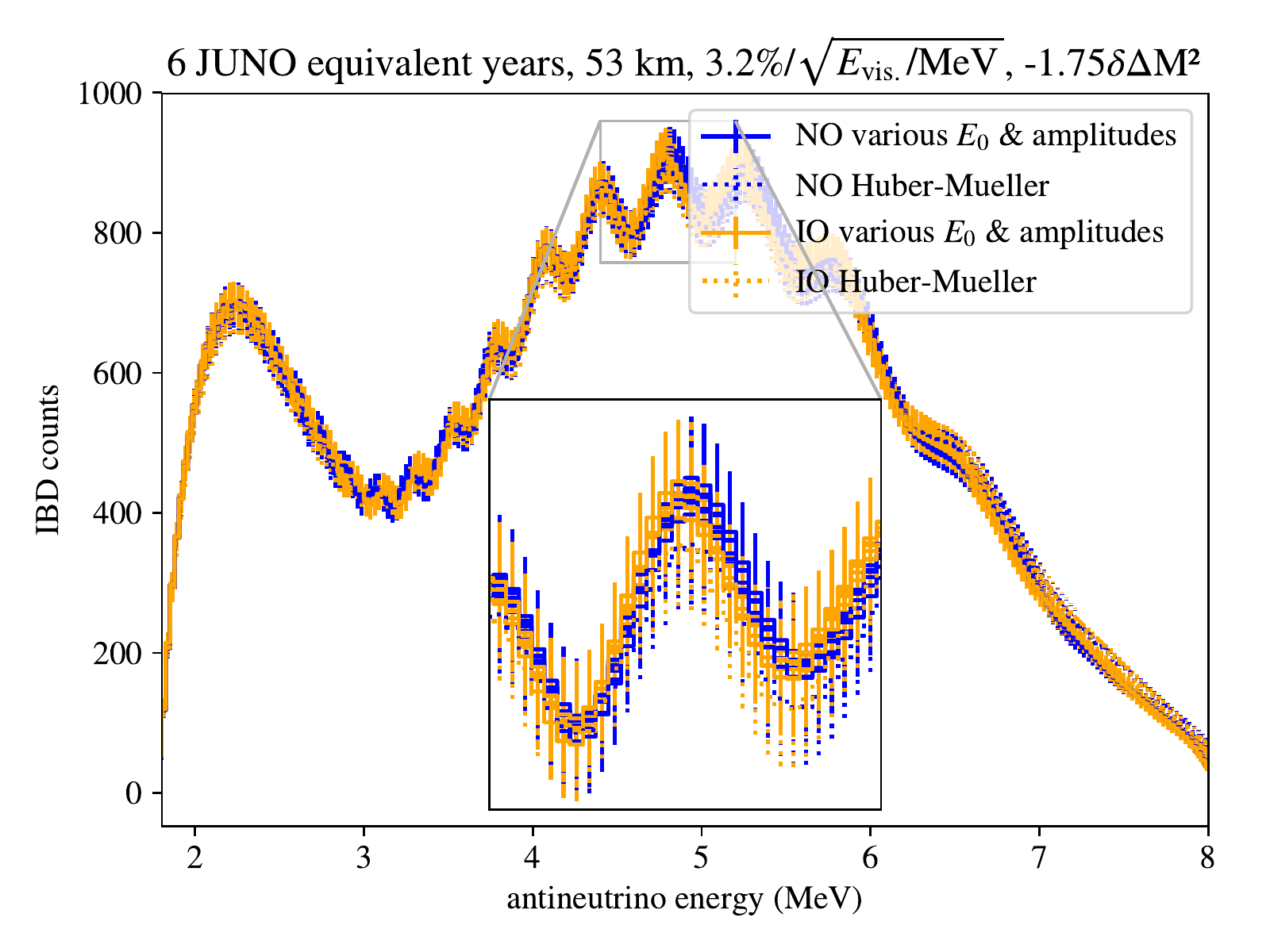}
\includegraphics[width= \linewidth]{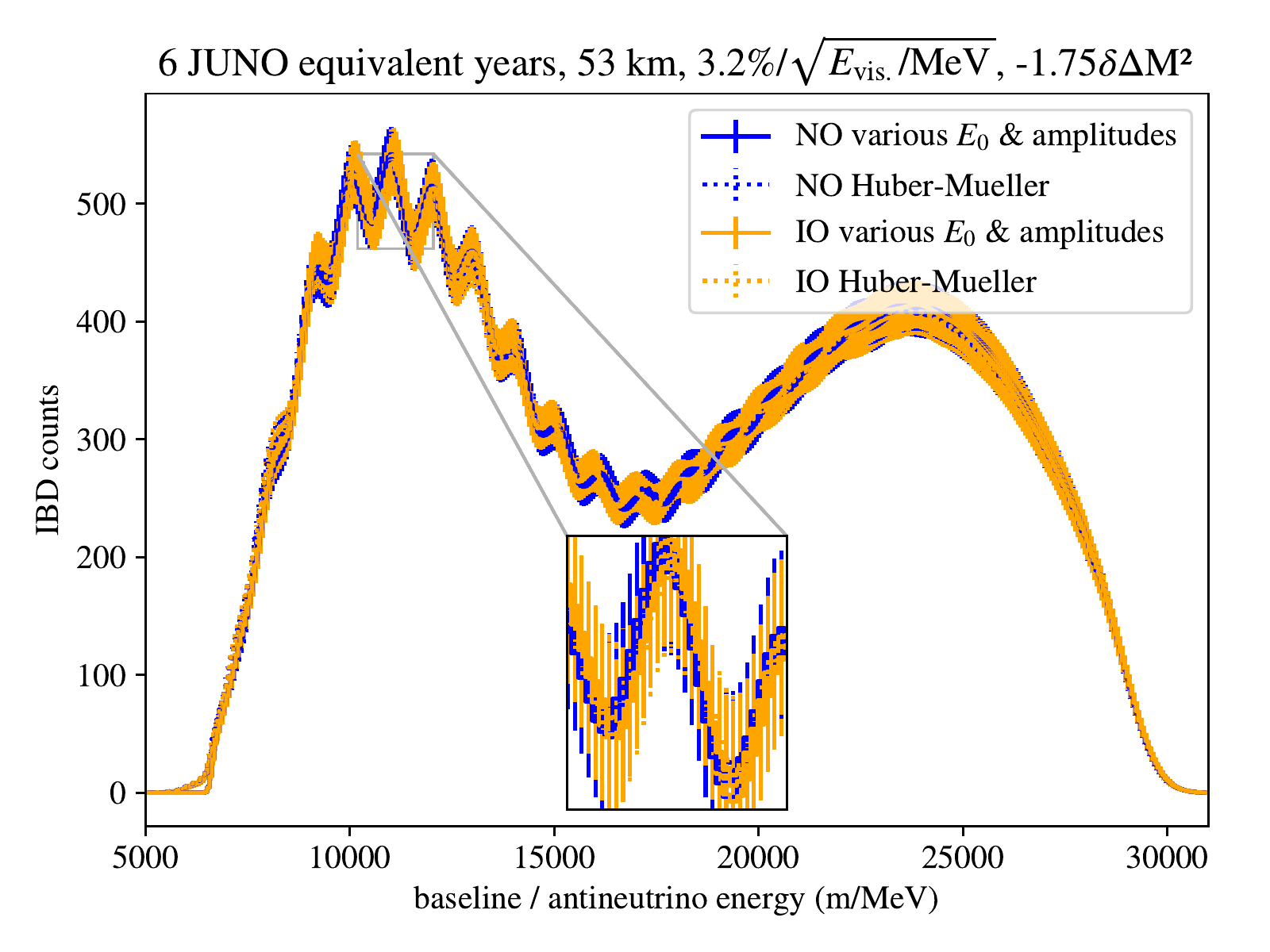}
\caption{The simulated result of six years' exposure given a JUNO-like rate of 60 per day inverse beta decays detected \cite{grassi}, for various plausible spectra give nearly degenerate hierarchy parameters, where the range of fine structures in the reactor spectra shown overwhelms the difference between the opposite mass hierarchies. The spectra were obtained by sampling branching ratios (amplitudes) and endpoint energies ($E_0$) as in Figure~\ref{fig:varied} for a detector given $3.2\%/\sqrt{E_\mathrm{vis.}/\mathrm{MeV}}$ at 53 km from the reactor. The oscillation parameter $\Delta M^2$ is shifted through -1.75 standard deviations from its central value for each hierarchy, while the remaining oscillation parameters retain their central values as given in Figure~\ref{fig:spectra}. Statistical error bars overwhelm the difference between the hierarchies' spectra for any given energy bin, and exacerbate the existing near degeneracy from the uncertain fine structure of the reactor spectra. However, despite the combined fine-structure and statistical uncertainties, this degeneracy is resolved by the Fourier cosine and sine transforms in Figure~\ref{fig:degeneracyFourier}.}
\label{fig:degeneracyData}
\end{figure}
As others have reported \cite{qian}, for certain natural values of normal- and inverted-ordering oscillation parameters, significant degeneracies conflate the observable oscillation patterns between the two orderings, posing a serious potential challenge to mass hierarchy experiments using reactor antineutrinos. The energy domain over which these degeneracies hold is partially determined by the achievable detector energy resolution, and partially determined by the distance from the reactor to the detector \cite{qian}. 

%\begin{figure}
%\includegraphics[width= \linewidth]{degeneracySpectra_a.pdf}
%\includegraphics[width= \linewidth]{degeneracySpectra_b.pdf}
%\caption{A sampling of plausible spectra for nearly degenerate hierarchy parameters, where the range of possible fine structures in the reactor spectra shown overwhelms the difference between the opposite mass hierarchies' spectra. The fine structure variations were obtained by sampling branching ratios (amplitudes) and endpoint energies ($E_0$) as in Figure~\ref{fig:varied}, and the spectra are plotted as a function of antineutrino energy on the left and $L/E$ on the right, for a detector given $3.2\%/\sqrt{E_\mathrm{vis.}/\mathrm{MeV}}$ 53 km from the reactor. The oscillation parameter $\Delta M^2$ is shifted through -1.75 standard deviations from its central value for each hierarchy, while the remaining oscillation parameters retain their central values as given in Figure~\ref{fig:spectra}. As a result the spectra appear degenerate in energy space due to the uncertain fine structure of the reactor spectra, but this degeneracy is resolved by the Fourier cosine and sine transforms in Figure~\ref{fig:degeneracyFourier}.}
%\label{fig:degeneracySpectra}
%\end{figure}

Figure~\ref{fig:degeneracyData} shows a nearly degenerate case with plausible hierarchy parameters after a simulated six year exposure, given a JUNO-like rate of 60 per day inverse beta decays detected \cite{grassi}. The hierarchies are rendered apparently degenerate in energy space by considering the various possible fine structures in the reactor spectra from Section \ref{sec:decayUncertainties}, whose range overwhelms the small difference between the opposite mass hierarchies' spectra, even before considering statistical uncertainties. This corresponds to a situation wherein one of these spectra is chosen by nature, but an opposite hierarchy spectrum forms a fitted hypothesis in near agreement, but wrong. Capozzi, Lisi, and Marrone have demonstrated that, in energy space, such nearly degenerate cases are not unreasonable to expect \cite{lisi}, and dominant statistical uncertainties further confound hierarchy determination in energy space.

Considering instead the Fourier cosine and sine transforms, however, the uncertainty in the fine structure of the reactor spectrum does not significantly impair hierarchy sensitivity even in the case of such near degeneracy. To see this requires statistical error propagation into the frequency domain, following a method adapted from reference~\cite{fourierError} while ensuring consistency between the uncertainty normalization and that of the transforms themselves:

\begin{equation}\label{eq:fourierError}
\sigma(\omega) = 2 \Delta\frac{L}{E}\sqrt{\sum_i g^2\left(\omega \left[\frac{L}{E}\right]_i\right) \sigma\left({\left[\frac{L}{E}\right]_i}\right)^2}
\end{equation}
where $g^2$ denotes $\sin^2$ or $\cos^2$ according to the coincident transform.

\begin{figure}
\includegraphics[width= \linewidth]{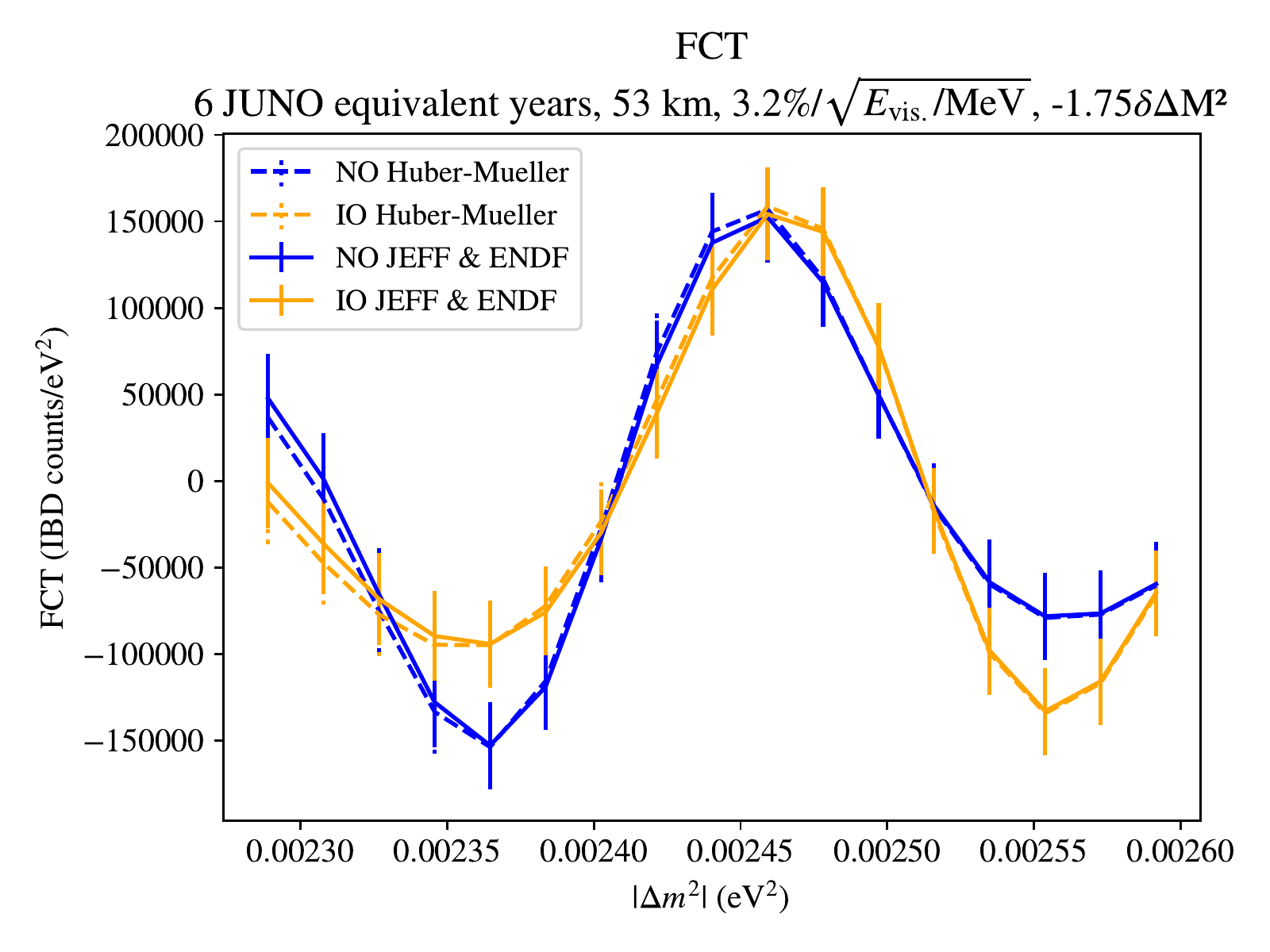}
\includegraphics[width= \linewidth]{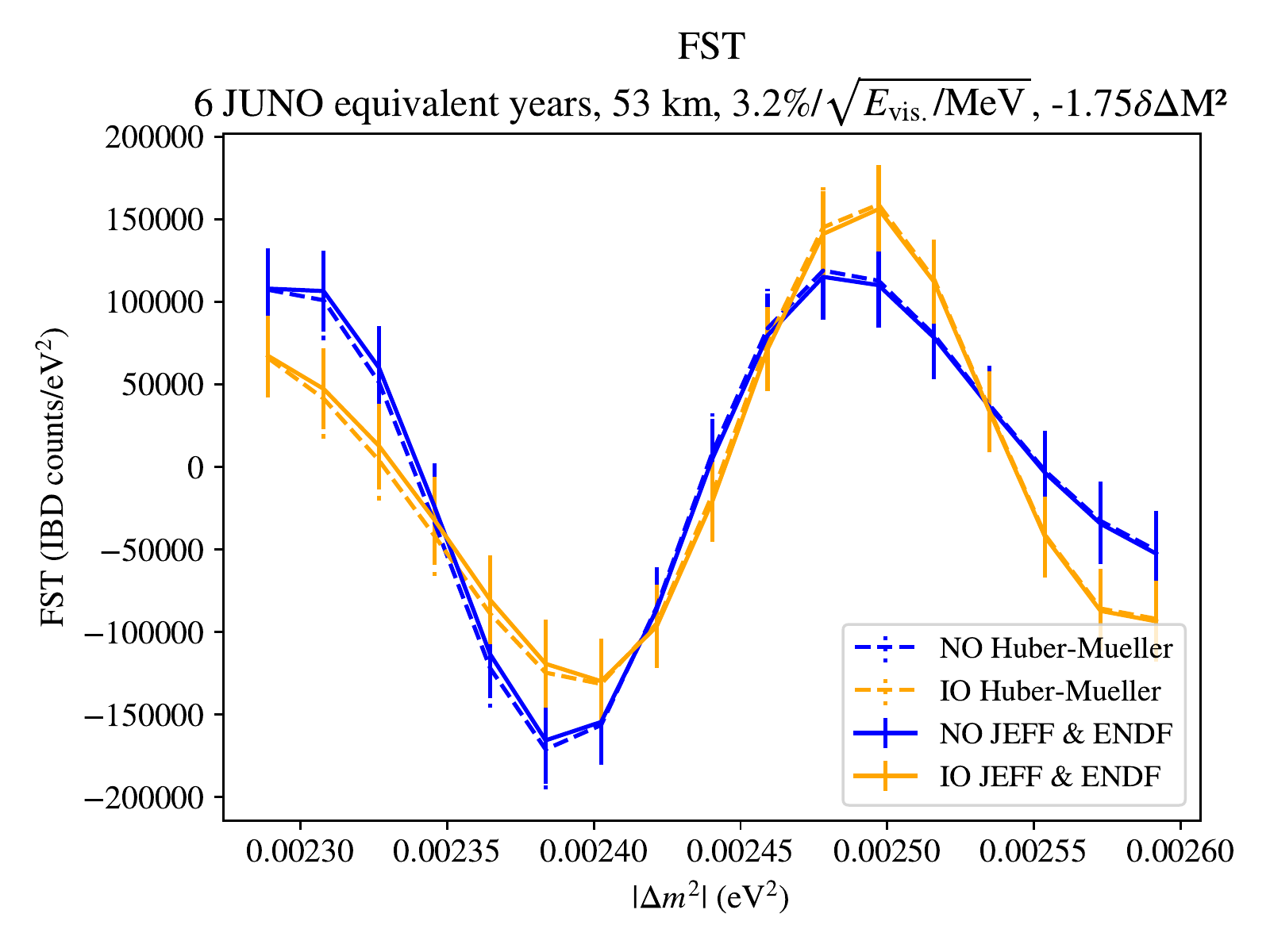}
\caption{The asymmetrical side-lobes of the Fourier cosine (FCT) and sine (FST) transforms distinguish the mass hierarchies even in the seemingly degenerate case from Figure~\ref{fig:degeneracyData}. The JEFF \& ENDF error bars comprise both systematic uncertainty in the fine structure of the reactor spectrum, and statistical uncertainty after six JUNO-equivalent years \cite{grassi}. The Fourier transforms resolve the degeneracy, as the difference between the opposite hierarchies' curves is recovered as the dominant effect beyond the range of the fine structure variations, due to the significant accumulation of phase and frequency correlations in the hierarchy-dependent oscillation patterns.}
\label{fig:degeneracyFourier}
\end{figure}

Figure~\ref{fig:degeneracyFourier} shows that the apparent degeneracy due to uncertainties in the reactor spectrum is resolved in Fourier space, where the difference between hierarchies is recovered as the dominant effect. This is again due to the fact that for the sawtooth contribution to significantly distort the hierarchy-dependent features in Fourier space would require the spacing of beta decay endpoint energies to precisely align with the spacing of hierarchy-dependent oscillation peaks throughout the energy domain. The chance of such alignment can be neglected, since the two systems are totally independent, and the endpoint energies and oscillation parameters are sufficiently constrained by experimental knowledge to rule out such a coincidence.

Although the uncertainties in the JEFF \& ENDF spectrum are not well known in general, a conservative error margin can be obtained within the hierarchy-sensitive region of Fourier space by noting that, with beta decay $Q$-values held fixed, even very large and diverse variations of the fine structure in the reactor spectrum all transform into a specific narrow band of curves in the pertinent frequency region. Having produced such varied spectra, we therefore quantify the fine-structure uncertainty as the maximum excursion over each frequency bin in the transform at that frequency, as shown in Figure~\ref{fig:degeneracyFourier}. Since the maximum excursion among neighboring bins often originates from some particular set of decay parameters, this uncertainty is treated as bin-to-bin correlated as a conservative choice.

\begin{figure}
\includegraphics[width=\linewidth]{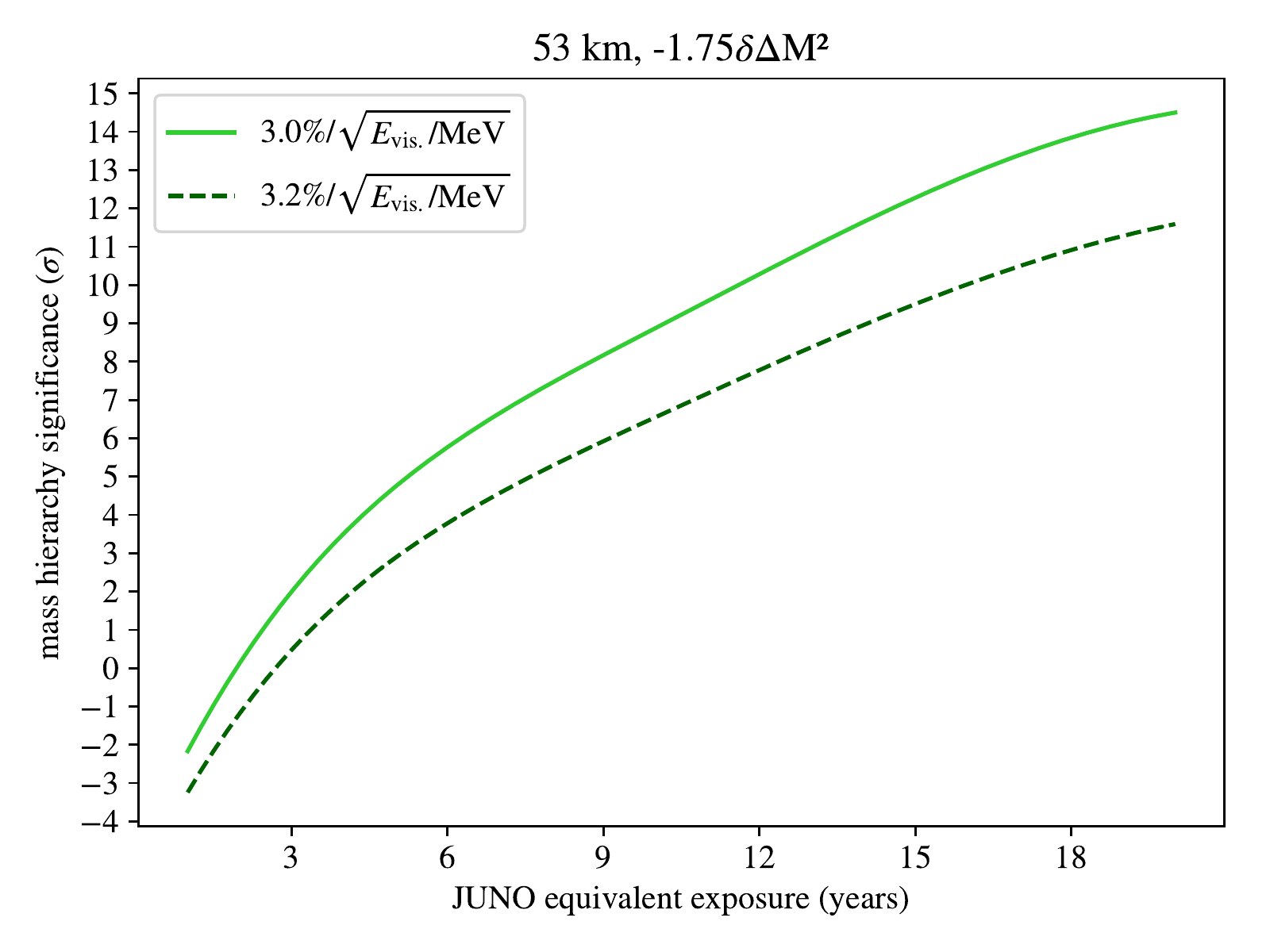}
\caption{Assuming the same nearly degenerate hierarchy parameters as Figure~\ref{fig:degeneracyData}, the predicted significance over time to rightly determine either mass hierarchy using Fourier analysis when considering fine-structure systematic uncertainty and statistical uncertainty. A JUNO-like rate of 60 per day inverse beta decays detected is assumed, in a detector 53~km from the reactor. The correct mass hierarchy is determined with $3.8\sigma$ ($5.8\sigma$) significance after six years given $3.2\%\:(3.0\%)/\sqrt{E_\mathrm{vis.}/\mathrm{MeV}}$ energy resolution. In a real experiment, other uncertainties will compound to reduce these significances. Nevertheless, since JUNO aims to achieve 3$\sigma$--4$\sigma$ significance within six years with $3.0\%/\sqrt{E_\mathrm{vis.}/\mathrm{MeV}}$ energy resolution, the uncertainty in the fine structure of the reactor spectrum in particular does not impede hierarchy determination in such experiments.}
\label{fig:degeneracySignificance}
\end{figure}

Figure~\ref{fig:degeneracySignificance} shows the significance to rightly determine either mass hierarchy over time given a JUNO-like IBD detection rate \cite{grassi}, for the same nearly degenerate case, and considering only the systematic uncertainty due to the fine structure of the reactor spectrum, combined with statistics. This significance derives from a $\chi^2$ test statistic evaluated over the hierarchy-sensitive region $0.00230 \le |\Delta m^2|/\mathrm{eV}^2 \le 0.00258$ in the combined Fourier cosine and sine transforms. When considering this combination of uncertainties alone, the mass hierarchy is detectable with $3.8\sigma$ ($5.8\sigma$) significance after six years of observation with $3.2\%\: (3.0\%)/\sqrt{E_\mathrm{vis.}/\mathrm{MeV}}$ energy resolution. With energy resolutions poorer than $3.2\%/\sqrt{E_\mathrm{vis.}/\mathrm{MeV}}$, this nearly-degenerate case is difficult to resolve even with perfect knowledge of the reactor spectrum. Given JUNO aims to achieve 3--4$\sigma$ significance after six years with $3.0\%/\sqrt{E_\mathrm{vis.}/\mathrm{MeV}}$ resolution \cite{JUNO}, the fine-structure uncertainty in the reactor spectrum is not a significant impediment to hierarchy determination with such an experiment.

As this example shows, however, the possibility of energy spectra that are nearly degenerate relative to the fine-structure uncertainty raises the critical importance of accurate Fourier analysis. In light of this, the following reviews the critical importance of good knowledge of the detector energy response.

\subsection{Nonlinearities in the Energy Response}\label{sec:nonlinearities}
As others have pointed out \cite{qian, lisi, dayabay2}, energy scale nonlinearities as small as one percent of the true energy can destroy the critical phase and frequency differences between the mass hierarchies' oscillation patterns. Thus any experiment aiming to distinguish a mass hierarchy using reactor antineutrinos must achieve unprecedented control of energy response nonlinearities \cite{qian}. If such nonlinearities are not sufficiently controlled, the resulting phase and frequency distortions can severely impede the hierarchy resolving power of the Fourier cosine and sine transforms \cite{qian}, whose success hinges mostly on the summation of accurate phase and frequency correlations over the entire $L/E$ domain.

As shown in the preceding section, the Fourier transforms mitigate the effects of the fine-structure uncertainties of the reactor spectrum even if the natural oscillation parameters render the energy spectra themselves degenerate within these uncertainties. Hence energy response nonlinearities must be controlled to below 0.5\%, to preserve the hierarchy resolving power of Fourier analysis \cite{dayabay2, qian}.

\section{Conclusion}\label{sec:conclusion}
\subsection{Overall Fourier Effect of the\\Spectral Distortions} \label{sec:zoomedOut}

\begin{figure}
\includegraphics[width= \linewidth]{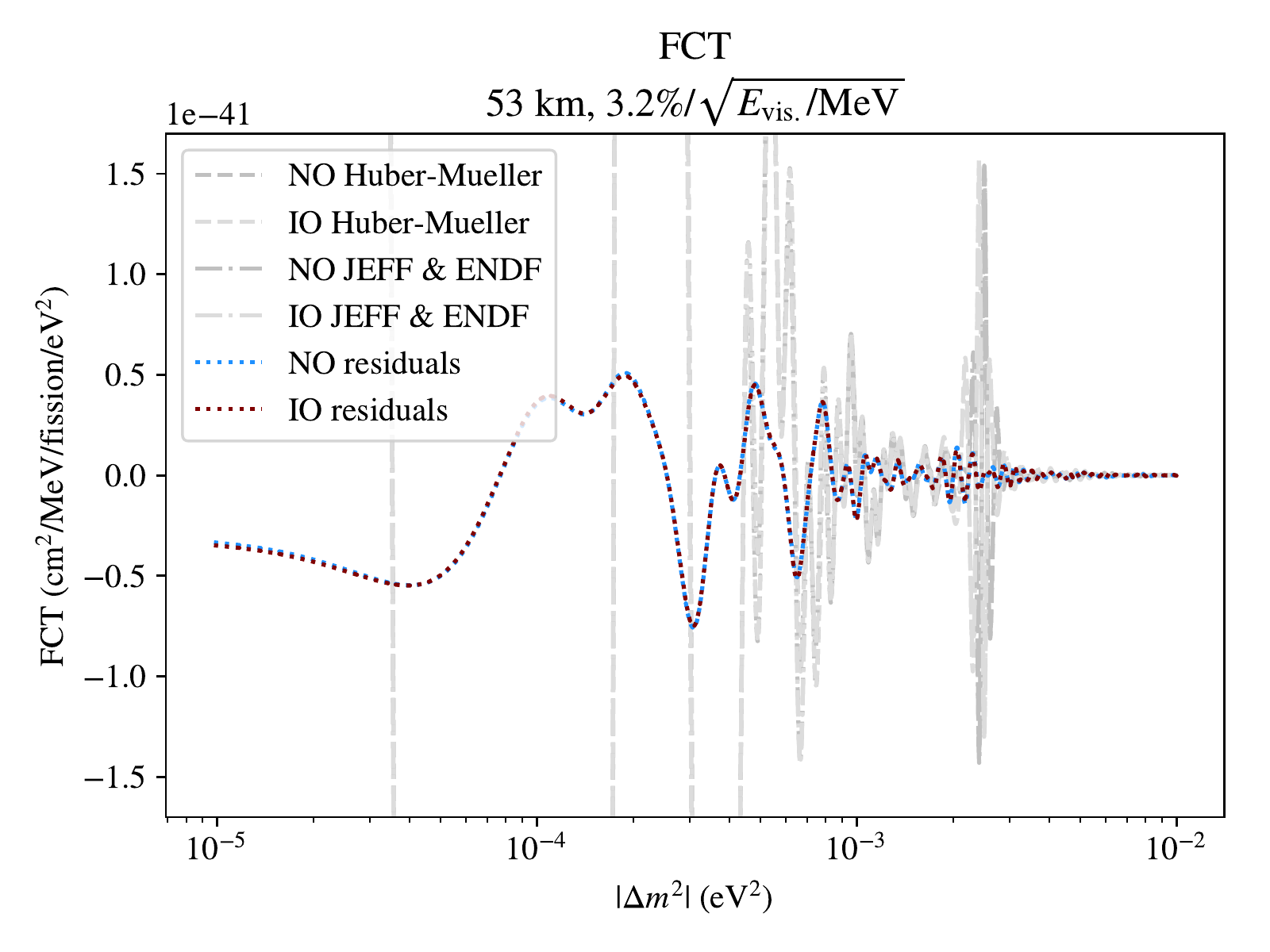} % Zoomed out transform
\includegraphics[width= \linewidth]{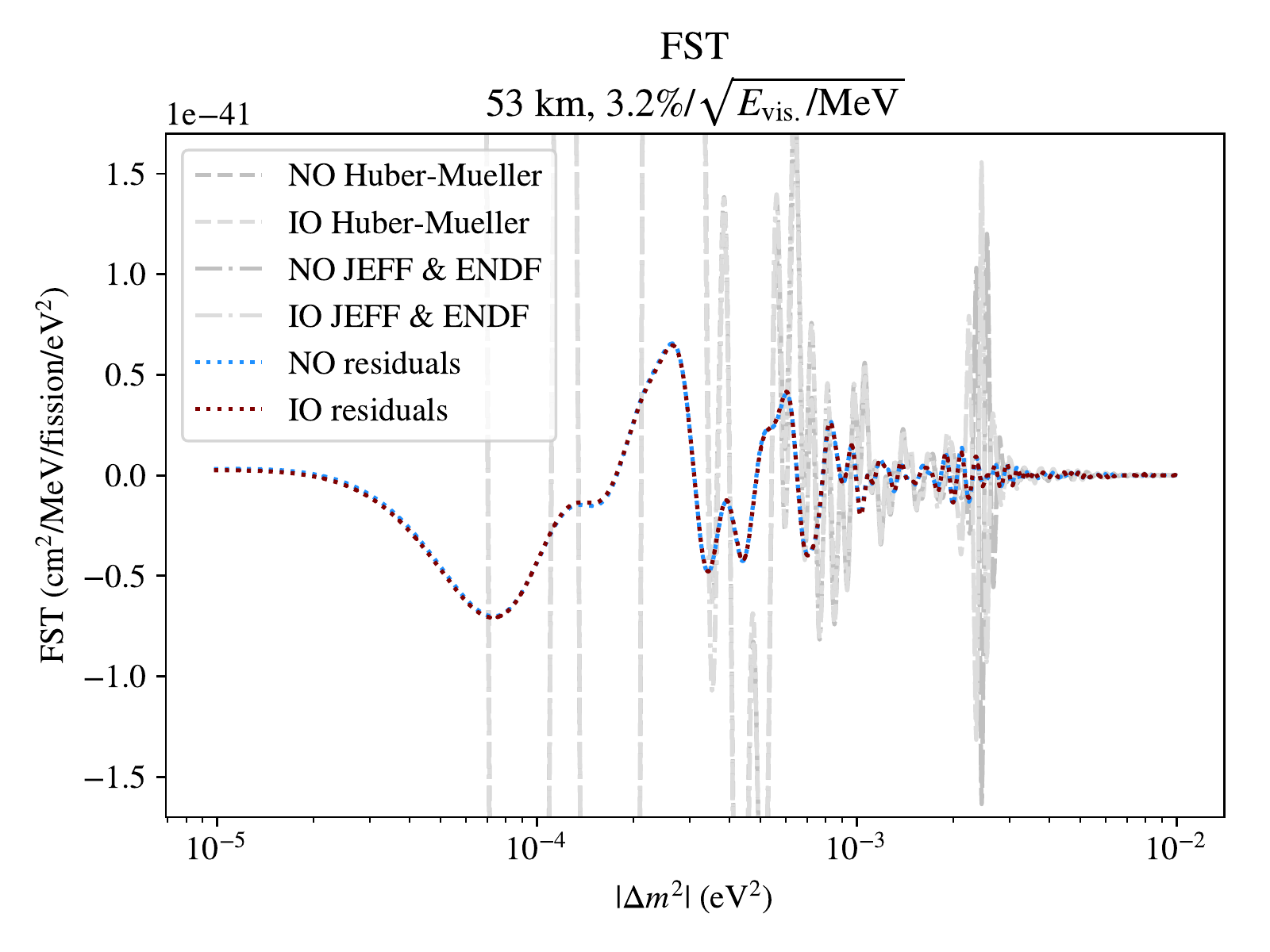} % Zoomed out transform
\caption{The Fourier cosine (FCT) and sine (FST) transforms of the JEFF~\&~ENDF and Huber-Mueller predicted antineutrino $L/E$ spectra, and the residuals between them, across a wide range of apparent $|\Delta m^2|$ oscillation frequencies from $10^{-5}\mathrm{eV}^2$ through $10^{-2}\mathrm{eV}^2$, for both normal ordered (NO) and inverted ordered (IO) neutrino mass hierarchies. A detector with $3.2\%/\sqrt{E_\mathrm{vis.}/\mathrm{MeV}}$ energy resolution at 53~km from the reactor is assumed, and the oscillation parameters were taken to be their central values as given in Figure~\ref{fig:spectra}.
	The overall contribution of the sawtooth effect to the Fourier transforms is small compared to neutrino oscillation effects throughout the domain of apparent oscillation frequencies, and concentrates in the decade from $10^{-4}\mathrm{eV}^2$ to $10^{-3}\mathrm{eV}^2$. The mass hierarchy-dependent features are prominent in the superordinate decade, in the region from $2.3\times10^{-3}\mathrm{eV}^2$ through $2.6\times10^{-3}\mathrm{eV}^2$, where the sawtooth contribution constitutes only a few percent relative to the $\Delta m^2_{31}$ peak as shown in detail in Figure~\ref{fig:smeared}.
}
\label{fig:zoomedOut}
\end{figure}
We have shown that in the frequency region around $\Delta m^2_{31}$, the magnitude of the effect from the sawtooth-like reactor-spectral features on the expected response of an antineutrino detector stands below an order of magnitude subordinate to the features critical to determining the neutrinos' mass ordering from their oscillation pattern. This led us to ask: what effects do the sawtooth-like features introduce when considering a broader domain in frequency space? Figure~\ref{fig:zoomedOut} shows the effect of the sawtooth features over a domain of apparent $|\Delta m^2|$ frequencies from $10^{-5}\mathrm{eV}^2$ through $10^{-2}\mathrm{eV}^2$. Above $10^{-3}\mathrm{eV}^2$ the sawtooth-like features originate no Fourier-spectral deviations that are of a magnitude similar to those features coming from the antineutrino oscillations themselves.

\subsection{Summary}\label{sec:summary}
It is important to understand the effects of the fine structure in a reactor antineutrino energy spectrum before attempting to resolve a neutrino mass hierarchy from oscillatory modulations of that spectrum. To that end, we have used the reactor antineutrino spectrum produced by JEFF~\&~ENDF to sum the constituent branch-weighted beta decay spectra in proportion to their cumulative fission yields. In doing so, we have simulated the abrupt Fermi-function cutoff features in the spectrum.

We find these spectral distortions and their uncertainties are uniformly negligible in magnitude throughout and well beyond the relevant oscillation frequency domain under Fourier analysis, and therefore they do not pose a challenge to neutrino mass hierarchy measurements using reactor antineutrinos. This result is robust under varied assumptions on the underlying beta decay endpoint energies and amplitudes. This result is also similarly robust in the face of realistic detector energy resolution. Moreover, this result holds throughout the experimentally allowed region in hierarchy parameter space, though substantial experimental challenges may yet arise given certain combinations of oscillation parameters \cite{qian}, independent of the structure of the underlying reactor spectrum which this work has addressed.

\begin{acknowledgments}
We thank Eligio Lisi for bringing the concerns addressed here to our attention, and for encouraging this investigation into them.
\end{acknowledgments}

\bibliography{JUNO-sawtooth}

\begin{thebibliography}{21}%
\makeatletter
\providecommand \@ifxundefined [1]{%
 \@ifx{#1\undefined}
}%
\providecommand \@ifnum [1]{%
 \ifnum #1\expandafter \@firstoftwo
 \else \expandafter \@secondoftwo
 \fi
}%
\providecommand \@ifx [1]{%
 \ifx #1\expandafter \@firstoftwo
 \else \expandafter \@secondoftwo
 \fi
}%
\providecommand \natexlab [1]{#1}%
\providecommand \enquote  [1]{``#1''}%
\providecommand \bibnamefont  [1]{#1}%
\providecommand \bibfnamefont [1]{#1}%
\providecommand \citenamefont [1]{#1}%
\providecommand \href@noop [0]{\@secondoftwo}%
\providecommand \href[0]{\begingroup \@sanitize@url \@href}%
\providecommand \@href[1]{\@@startlink{#1}\@@href}%
\providecommand \@@href[1]{\endgroup#1\@@endlink}%
\providecommand \@sanitize@url [0]{\catcode `\\12\catcode `\$12\catcode
  `\&12\catcode `\#12\catcode `\^12\catcode `\_12\catcode `\%12\relax}%
\providecommand \@@startlink[1]{}%
\providecommand \@@endlink[0]{}%
\providecommand \url  [0]{\begingroup\@sanitize@url \@url }%
\providecommand \@url [1]{\endgroup\@href {#1}{\urlprefix }}%
\providecommand \urlprefix  [0]{URL }%
\providecommand \Eprint [0]{\href }%
\providecommand \doibase [0]{http://dx.doi.org/}%
\providecommand \selectlanguage [0]{\@gobble}%
\providecommand \bibinfo  [0]{\@secondoftwo}%
\providecommand \bibfield  [0]{\@secondoftwo}%
\providecommand \translation [1]{[#1]}%
\providecommand \BibitemOpen [0]{}%
\providecommand \bibitemStop [0]{}%
\providecommand \bibitemNoStop [0]{.\EOS\space}%
\providecommand \EOS [0]{\spacefactor3000\relax}%
\providecommand \BibitemShut  [1]{\csname bibitem#1\endcsname}%
\let\auto@bib@innerbib\@empty
%</preamble>
\bibitem [{\citenamefont {An}\ \emph {et~al.}(2016)\citenamefont {An},
  \citenamefont {An}, \citenamefont {An}, \citenamefont {Antonelli},
  \citenamefont {Baussan}, \citenamefont {Beacom}, \citenamefont {Bezrukov},
  \citenamefont {Blyth}, \citenamefont {Brugnera}, \citenamefont {Avanzini},\
  and\ \citenamefont {et~al.}}]{JUNO}%
  \BibitemOpen
  \bibfield  {author} {\bibinfo {author} {\bibfnamefont {F.}~\bibnamefont
  {An}}, \bibinfo {author} {\bibfnamefont {G.}~\bibnamefont {An}}, \bibinfo
  {author} {\bibfnamefont {Q.}~\bibnamefont {An}}, \bibinfo {author}
  {\bibfnamefont {V.}~\bibnamefont {Antonelli}},  \emph {et~al.},\ }\enquote
  {\bibinfo {title} {Neutrino physics with JUNO},}\ \href{\doibase
  10.1088/0954-3899/43/3/030401} {\bibfield  {journal} {\bibinfo  {journal}
  {Journal of Physics G: Nuclear and Particle Physics}\ }\textbf {\bibinfo
  {volume} {43}},\ \bibinfo {pages} {030401} (\bibinfo {year}
  {2016})}\BibitemShut {NoStop}%
\bibitem [{\citenamefont {Petcov}\ and\ \citenamefont {Piai}(2002)}]{Petcov}%
  \BibitemOpen
  \bibfield  {author} {\bibinfo {author} {\bibfnamefont {S.}~\bibnamefont
  {Petcov}}\ and\ \bibinfo {author} {\bibfnamefont {M.}~\bibnamefont {Piai}},\
  }\enquote {\bibinfo {title} {The LMA MSW solution of the solar neutrino
  problem, inverted neutrino mass hierarchy and reactor neutrino
  experiments},}\ \href{\doibase 10.1016/s0370-2693(02)01591-5} {\bibfield
  {journal} {\bibinfo  {journal} {Physics Letters B}\ }\textbf {\bibinfo
  {volume} {533}},\ \bibinfo {pages} {94} (\bibinfo {year} {2002})}\BibitemShut
  {NoStop}%
\bibitem [{\citenamefont {Capozzi}\ \emph {et~al.}(2015)\citenamefont
  {Capozzi}, \citenamefont {Lisi},\ and\ \citenamefont {Marrone}}]{lisi}%
  \BibitemOpen
  \bibfield  {author} {\bibinfo {author} {\bibfnamefont {F.}~\bibnamefont
  {Capozzi}}, \bibinfo {author} {\bibfnamefont {E.}~\bibnamefont {Lisi}}, \
  and\ \bibinfo {author} {\bibfnamefont {A.}~\bibnamefont {Marrone}},\
  }\enquote {\bibinfo {title} {Neutrino mass hierarchy and precision physics
  with medium-baseline reactors: Impact of energy-scale and flux-shape
  uncertainties},}\ \href{\doibase 10.1103/PhysRevD.92.093011} {\bibfield
  {journal} {\bibinfo  {journal} {Phys. Rev. D}\ }\textbf {\bibinfo {volume}
  {92}},\ \bibinfo {pages} {093011} (\bibinfo {year} {2015})}\BibitemShut
  {NoStop}%
\bibitem [{\citenamefont {Qian}\ \emph {et~al.}(2013)\citenamefont {Qian},
  \citenamefont {Dwyer}, \citenamefont {McKeown}, \citenamefont {Vogel},
  \citenamefont {Wang},\ and\ \citenamefont {Zhang}}]{qian}%
  \BibitemOpen
  \bibfield  {author} {\bibinfo {author} {\bibfnamefont {X.}~\bibnamefont
  {Qian}}, \bibinfo {author} {\bibfnamefont {D.~A.}\ \bibnamefont {Dwyer}},
  \bibinfo {author} {\bibfnamefont {R.~D.}\ \bibnamefont {McKeown}}, \bibinfo
  {author} {\bibfnamefont {P.}~\bibnamefont {Vogel}}, \bibinfo {author}
  {\bibfnamefont {W.}~\bibnamefont {Wang}}, \ and\ \bibinfo {author}
  {\bibfnamefont {C.}~\bibnamefont {Zhang}},\ }\enquote {\bibinfo {title} {Mass
  hierarchy resolution in reactor anti-neutrino experiments: Parameter
  degeneracies and detector energy response},}\ \href{\doibase
  10.1103/PhysRevD.87.033005} {\bibfield  {journal} {\bibinfo  {journal} {Phys.
  Rev. D}\ }\textbf {\bibinfo {volume} {87}},\ \bibinfo {pages} {033005}
  (\bibinfo {year} {2013})}\BibitemShut {NoStop}%
\bibitem [{\citenamefont {Forero}\ \emph {et~al.}(2017)\citenamefont {Forero},
  \citenamefont {Hawkins},\ and\ \citenamefont {Huber}}]{forero}%
  \BibitemOpen
  \bibfield  {author} {\bibinfo {author} {\bibfnamefont {D.~V.}\ \bibnamefont
  {Forero}}, \bibinfo {author} {\bibfnamefont {R.}~\bibnamefont {Hawkins}}, \
  and\ \bibinfo {author} {\bibfnamefont {P.}~\bibnamefont {Huber}},\ }\enquote
  {\bibinfo {title} {The benefits of a near detector for JUNO},}\ \href@noop {}
  {\  (\bibinfo {year} {2017})},\ \Eprint {http://arxiv.org/abs/1710.07378}
  {arXiv:1710.07378 [hep-ph]} \BibitemShut {NoStop}%
%%CITATION = ARXIV:1710.07378;%%
\bibitem [{\citenamefont {de~Salas}\ \emph {et~al.}(2018)\citenamefont
  {de~Salas}, \citenamefont {Forero}, \citenamefont {Ternes}, \citenamefont
  {T{\'o}rtola},\ and\ \citenamefont {Valle}}]{deSalas}%
  \BibitemOpen
  \bibfield  {author} {\bibinfo {author} {\bibfnamefont {P.}~\bibnamefont
  {de~Salas}}, \bibinfo {author} {\bibfnamefont {D.}~\bibnamefont {Forero}},
  \bibinfo {author} {\bibfnamefont {C.}~\bibnamefont {Ternes}}, \bibinfo
  {author} {\bibfnamefont {M.}~\bibnamefont {T{\'o}rtola}}, \ and\ \bibinfo
  {author} {\bibfnamefont {J.}~\bibnamefont {Valle}},\ }\enquote {\bibinfo
  {title} {Status of neutrino oscillations 2018: 3$\sigma$ hint for normal mass
  ordering and improved CP sensitivity},}\ \href{\doibase
  https://doi.org/10.1016/j.physletb.2018.06.019} {\bibfield  {journal}
  {\bibinfo  {journal} {Physics Letters B}\ }\textbf {\bibinfo {volume}
  {782}},\ \bibinfo {pages} {633 } (\bibinfo {year} {2018})}\BibitemShut
  {NoStop}%
\bibitem [{\citenamefont {Hayes}\ \emph {et~al.}(2015)\citenamefont {Hayes},
  \citenamefont {Friar}, \citenamefont {Garvey}, \citenamefont {Ibeling},
  \citenamefont {Jungman}, \citenamefont {Kawano},\ and\ \citenamefont
  {Mills}}]{Hayes-Bump}%
  \BibitemOpen
  \bibfield  {author} {\bibinfo {author} {\bibfnamefont {A.~C.}\ \bibnamefont
  {Hayes}}, \bibinfo {author} {\bibfnamefont {J.~L.}\ \bibnamefont {Friar}},
  \bibinfo {author} {\bibfnamefont {G.~T.}\ \bibnamefont {Garvey}}, \bibinfo
  {author} {\bibfnamefont {D.}~\bibnamefont {Ibeling}}, \bibinfo {author}
  {\bibfnamefont {G.}~\bibnamefont {Jungman}}, \bibinfo {author} {\bibfnamefont
  {T.}~\bibnamefont {Kawano}}, \ and\ \bibinfo {author} {\bibfnamefont {R.~W.}\
  \bibnamefont {Mills}},\ }\enquote {\bibinfo {title} {Possible origins and
  implications of the shoulder in reactor neutrino spectra},}\ \href{\doibase
  10.1103/PhysRevD.92.033015} {\bibfield  {journal} {\bibinfo  {journal} {Phys.
  Rev. D}\ }\textbf {\bibinfo {volume} {92}},\ \bibinfo {pages} {033015}
  (\bibinfo {year} {2015})}\BibitemShut {NoStop}%
\bibitem [{\citenamefont {Sonzogni}\ \emph {et~al.}(2018)\citenamefont
  {Sonzogni}, \citenamefont {Nino},\ and\ \citenamefont
  {McCutchan}}]{fine-structure}%
  \BibitemOpen
  \bibfield  {author} {\bibinfo {author} {\bibfnamefont {A.~A.}\ \bibnamefont
  {Sonzogni}}, \bibinfo {author} {\bibfnamefont {M.}~\bibnamefont {Nino}}, \
  and\ \bibinfo {author} {\bibfnamefont {E.~A.}\ \bibnamefont {McCutchan}},\
  }\enquote {\bibinfo {title} {Revealing fine structure in the antineutrino
  spectra from a nuclear reactor},}\ \href{\doibase 10.1103/PhysRevC.98.014323}
  {\bibfield  {journal} {\bibinfo  {journal} {Phys. Rev. C}\ }\textbf {\bibinfo
  {volume} {98}},\ \bibinfo {pages} {014323} (\bibinfo {year}
  {2018})}\BibitemShut {NoStop}%
\bibitem [{\citenamefont {An}\ \emph {et~al.}(2012)\citenamefont {An},
  \citenamefont {Bai}, \citenamefont {Balantekin}, \citenamefont {Band},
  \citenamefont {Beavis}, \citenamefont {Beriguete}, \citenamefont {Bishai},
  \citenamefont {Blyth}, \citenamefont {Boddy}, \citenamefont {Brown},
  \citenamefont {Cai}, \citenamefont {Cao}, \citenamefont {Cao}, \citenamefont
  {Carr}, \citenamefont {Chan}, \citenamefont {Chang}, \citenamefont {Chang},
  \citenamefont {Chasman}, \citenamefont {Chen}, \citenamefont {Chen},
  \citenamefont {Chen}, \citenamefont {Chen}, \citenamefont {Chen},
  \citenamefont {Chen}, \citenamefont {Chen}, \citenamefont {Chen},
  \citenamefont {Chen}, \citenamefont {Cherwinka}, \citenamefont {Chu},
  \citenamefont {Cummings}, \citenamefont {Deng}, \citenamefont {Ding},
  \citenamefont {Diwan}, \citenamefont {Dong}, \citenamefont {Draeger},
  \citenamefont {Du}, \citenamefont {Dwyer}, \citenamefont {Edwards},
  \citenamefont {Ely}, \citenamefont {Fang}, \citenamefont {Fu}, \citenamefont
  {Fu}, \citenamefont {Ge}, \citenamefont {Ghazikhanian}, \citenamefont {Gill},
  \citenamefont {Goett}, \citenamefont {Gonchar}, \citenamefont {Gong},
  \citenamefont {Gong}, \citenamefont {Gornushkin}, \citenamefont {Greenler},
  \citenamefont {Gu}, \citenamefont {Guan}, \citenamefont {Guo}, \citenamefont
  {Hackenburg}, \citenamefont {Hahn}, \citenamefont {Hans}, \citenamefont {He},
  \citenamefont {He}, \citenamefont {He}, \citenamefont {Heeger}, \citenamefont
  {Heng}, \citenamefont {Hinrichs}, \citenamefont {Ho}, \citenamefont {Hor},
  \citenamefont {Hsiung}, \citenamefont {Hu}, \citenamefont {Hu}, \citenamefont
  {Hu}, \citenamefont {Huang}, \citenamefont {Huang}, \citenamefont {Huang},
  \citenamefont {Huang}, \citenamefont {Huang}, \citenamefont {Huber},
  \citenamefont {Isvan}, \citenamefont {Jaffe}, \citenamefont {Jetter},
  \citenamefont {Ji}, \citenamefont {Ji}, \citenamefont {Jiang}, \citenamefont
  {Jiang}, \citenamefont {Jiao}, \citenamefont {Johnson}, \citenamefont {Kang},
  \citenamefont {Kettell}, \citenamefont {Kramer}, \citenamefont {Kwan},
  \citenamefont {Kwok}, \citenamefont {Kwok}, \citenamefont {Lai},
  \citenamefont {Lai}, \citenamefont {Lai}, \citenamefont {Lau}, \citenamefont
  {Lebanowski}, \citenamefont {Lee}, \citenamefont {Lee}, \citenamefont
  {Leitner}, \citenamefont {Leung}, \citenamefont {Leung}, \citenamefont
  {Lewis}, \citenamefont {Li}, \citenamefont {Li}, \citenamefont {Li},
  \citenamefont {Li}, \citenamefont {Li}, \citenamefont {Li}, \citenamefont
  {Li}, \citenamefont {Li}, \citenamefont {Li}, \citenamefont {Li},
  \citenamefont {Li}, \citenamefont {Li}, \citenamefont {Liang}, \citenamefont
  {Liang}, \citenamefont {Lin}, \citenamefont {Lin}, \citenamefont {Lin},
  \citenamefont {Lin}, \citenamefont {Lin}, \citenamefont {Ling}, \citenamefont
  {Link}, \citenamefont {Littenberg}, \citenamefont {Littlejohn}, \citenamefont
  {Liu}, \citenamefont {Liu}, \citenamefont {Liu}, \citenamefont {Liu},
  \citenamefont {Liu}, \citenamefont {Liu}, \citenamefont {Liu}, \citenamefont
  {Liu}, \citenamefont {Liu}, \citenamefont {Lu}, \citenamefont {Lu},
  \citenamefont {Luk}, \citenamefont {Luk}, \citenamefont {Luo}, \citenamefont
  {Luo}, \citenamefont {Ma}, \citenamefont {Ma}, \citenamefont {Ma},
  \citenamefont {Ma}, \citenamefont {Ma}, \citenamefont {Mayes}, \citenamefont
  {McDonald}, \citenamefont {McFarlane}, \citenamefont {McKeown}, \citenamefont
  {Meng}, \citenamefont {Mohapatra}, \citenamefont {Morgan}, \citenamefont
  {Nakajima}, \citenamefont {Napolitano}, \citenamefont {Naumov}, \citenamefont
  {Nemchenok}, \citenamefont {Newsom}, \citenamefont {Ngai}, \citenamefont
  {Ngai}, \citenamefont {Nie}, \citenamefont {Ning}, \citenamefont
  {Ochoa-Ricoux}, \citenamefont {Oh}, \citenamefont {Olshevski}, \citenamefont
  {Pagac}, \citenamefont {Patton}, \citenamefont {Pearson}, \citenamefont
  {Pec}, \citenamefont {Peng}, \citenamefont {Piilonen}, \citenamefont
  {Pinsky}, \citenamefont {Pun}, \citenamefont {Qi}, \citenamefont {Qi},
  \citenamefont {Qian}, \citenamefont {Raper}, \citenamefont {Rosero},
  \citenamefont {Roskovec}, \citenamefont {Ruan}, \citenamefont {Seilhan},
  \citenamefont {Shao}, \citenamefont {Shih}, \citenamefont {Steiner},
  \citenamefont {Stoler}, \citenamefont {Sun}, \citenamefont {Sun},
  \citenamefont {Tam}, \citenamefont {Tanaka}, \citenamefont {Tang},
  \citenamefont {Themann}, \citenamefont {Torun}, \citenamefont {Trentalange},
  \citenamefont {Tsai}, \citenamefont {Tsang}, \citenamefont {Tsang},
  \citenamefont {Tull}, \citenamefont {Viren}, \citenamefont {Virostek},
  \citenamefont {Vorobel}, \citenamefont {Wang}, \citenamefont {Wang},
  \citenamefont {Wang}, \citenamefont {Wang}, \citenamefont {Wang},
  \citenamefont {Wang}, \citenamefont {Wang}, \citenamefont {Wang},
  \citenamefont {Wang}, \citenamefont {Wang}, \citenamefont {Wang},
  \citenamefont {Wang}, \citenamefont {Wang}, \citenamefont {Wang},
  \citenamefont {Wang}, \citenamefont {Webber}, \citenamefont {Wei},
  \citenamefont {Wen}, \citenamefont {Wenman}, \citenamefont {Whisnant},
  \citenamefont {White}, \citenamefont {Whitehead}, \citenamefont {Whitten},
  \citenamefont {Wilhelmi}, \citenamefont {Wise}, \citenamefont {Wong},
  \citenamefont {Wong}, \citenamefont {Wong}, \citenamefont {Worcester},
  \citenamefont {Wu}, \citenamefont {Wu}, \citenamefont {Xia}, \citenamefont
  {Xiang}, \citenamefont {Xiao}, \citenamefont {Xing}, \citenamefont {Xu},
  \citenamefont {Xu}, \citenamefont {Xu}, \citenamefont {Xu}, \citenamefont
  {Xu}, \citenamefont {Xu}, \citenamefont {Xue}, \citenamefont {Yang},
  \citenamefont {Yang}, \citenamefont {Ye}, \citenamefont {Yeh}, \citenamefont
  {Yeh}, \citenamefont {Yip}, \citenamefont {Young}, \citenamefont {Yu},
  \citenamefont {Zhan}, \citenamefont {Zhang}, \citenamefont {Zhang},
  \citenamefont {Zhang}, \citenamefont {Zhang}, \citenamefont {Zhang},
  \citenamefont {Zhang}, \citenamefont {Zhang}, \citenamefont {Zhang},
  \citenamefont {Zhang}, \citenamefont {Zhang}, \citenamefont {Zhang},
  \citenamefont {Zhang}, \citenamefont {Zhang}, \citenamefont {Zhao},
  \citenamefont {Zhao}, \citenamefont {Zhao}, \citenamefont {Zheng},
  \citenamefont {Zhong}, \citenamefont {Zhou}, \citenamefont {Zhou},
  \citenamefont {Zhuang},\ and\ \citenamefont {Zou}}]{DayaBay}%
  \BibitemOpen
  \bibfield  {author} {\bibinfo {author} {\bibfnamefont {F.~P.}\ \bibnamefont
  {An}}, \bibinfo {author} {\bibfnamefont {J.~Z.}\ \bibnamefont {Bai}},
  \bibinfo {author} {\bibfnamefont {A.~B.}\ \bibnamefont {Balantekin}},
  \bibinfo {author} {\bibfnamefont {H.~R.}\ \bibnamefont {Band}},  \emph
  {et~al.},\ }\enquote {\bibinfo {title} {Observation of Electron-Antineutrino
  Disappearance at Daya Bay},}\ \href{\doibase 10.1103/PhysRevLett.108.171803}
  {\bibfield  {journal} {\bibinfo  {journal} {Phys. Rev. Lett.}\ }\textbf
  {\bibinfo {volume} {108}},\ \bibinfo {pages} {171803} (\bibinfo {year}
  {2012})}\BibitemShut {NoStop}%
\bibitem [{\citenamefont {Huber}(2011)}]{huber}%
  \BibitemOpen
  \bibfield  {author} {\bibinfo {author} {\bibfnamefont {P.}~\bibnamefont
  {Huber}},\ }\enquote {\bibinfo {title} {Determination of antineutrino spectra
  from nuclear reactors},}\ \href{\doibase 10.1103/PhysRevC.84.024617}
  {\bibfield  {journal} {\bibinfo  {journal} {Phys. Rev. C}\ }\textbf {\bibinfo
  {volume} {84}},\ \bibinfo {pages} {024617} (\bibinfo {year}
  {2011})}\BibitemShut {NoStop}%
\bibitem [{\citenamefont {Mueller}\ \emph {et~al.}(2011)\citenamefont
  {Mueller}, \citenamefont {Lhuillier}, \citenamefont {Fallot}, \citenamefont
  {Letourneau}, \citenamefont {Cormon}, \citenamefont {Fechner}, \citenamefont
  {Giot}, \citenamefont {Lasserre}, \citenamefont {Martino}, \citenamefont
  {Mention}, \citenamefont {Porta},\ and\ \citenamefont {Yermia}}]{mueller}%
  \BibitemOpen
  \bibfield  {author} {\bibinfo {author} {\bibfnamefont {T.~A.}\ \bibnamefont
  {Mueller}}, \bibinfo {author} {\bibfnamefont {D.}~\bibnamefont {Lhuillier}},
  \bibinfo {author} {\bibfnamefont {M.}~\bibnamefont {Fallot}}, \bibinfo
  {author} {\bibfnamefont {A.}~\bibnamefont {Letourneau}},  \emph {et~al.},\
  }\enquote {\bibinfo {title} {Improved predictions of reactor antineutrino
  spectra},}\ \href{\doibase 10.1103/PhysRevC.83.054615} {\bibfield  {journal}
  {\bibinfo  {journal} {Phys. Rev. C}\ }\textbf {\bibinfo {volume} {83}},\
  \bibinfo {pages} {054615} (\bibinfo {year} {2011})}\BibitemShut {NoStop}%
\bibitem [{\citenamefont {Li}\ \emph {et~al.}(2013)\citenamefont {Li},
  \citenamefont {Cao}, \citenamefont {Wang},\ and\ \citenamefont
  {Zhan}}]{dayabay2}%
  \BibitemOpen
  \bibfield  {author} {\bibinfo {author} {\bibfnamefont {Y.-F.}\ \bibnamefont
  {Li}}, \bibinfo {author} {\bibfnamefont {J.}~\bibnamefont {Cao}}, \bibinfo
  {author} {\bibfnamefont {Y.}~\bibnamefont {Wang}}, \ and\ \bibinfo {author}
  {\bibfnamefont {L.}~\bibnamefont {Zhan}},\ }\enquote {\bibinfo {title}
  {Unambiguous determination of the neutrino mass hierarchy using reactor
  neutrinos},}\ \href{\doibase 10.1103/PhysRevD.88.013008} {\bibfield
  {journal} {\bibinfo  {journal} {Phys. Rev. D}\ }\textbf {\bibinfo {volume}
  {88}},\ \bibinfo {pages} {013008} (\bibinfo {year} {2013})}\BibitemShut
  {NoStop}%
\bibitem [{\citenamefont {Kellett}\ \emph {et~al.}(2009)\citenamefont
  {Kellett}, \citenamefont {Bernillon},\ and\ \citenamefont {Mills}}]{JEFF}%
  \BibitemOpen
  \bibfield  {author} {\bibinfo {author} {\bibfnamefont {M.~A.}\ \bibnamefont
  {Kellett}}, \bibinfo {author} {\bibfnamefont {O.}~\bibnamefont {Bernillon}},
  \ and\ \bibinfo {author} {\bibfnamefont {R.~W.}\ \bibnamefont {Mills}},\
  }\href{https://www.oecd-nea.org/dbdata/nds_jefreports/jefreport-20/nea6287-jeff-20.pdf}
  {\emph {\bibinfo {title} {JEFF Report 20, The JEFF-3.1/-3.1.1 radioactive
  decay data and fission yields sub-libraries}}}\ (\bibinfo  {publisher}
  {Organisation for Economic Co-Operation and Development--Nuclear Energy
  Agency},\ \bibinfo {year} {2009})\BibitemShut {NoStop}%
\bibitem [{\citenamefont {Brown}\ \emph {et~al.}(2018)\citenamefont {Brown},
  \citenamefont {Chadwick}, \citenamefont {Capote}, \citenamefont {Kahler},
  \citenamefont {Trkov}, \citenamefont {Herman}, \citenamefont {Sonzogni},
  \citenamefont {Danon}, \citenamefont {Carlson}, \citenamefont {Dunn},
  \citenamefont {Smith}, \citenamefont {Hale}, \citenamefont {Arbanas},
  \citenamefont {Arcilla}, \citenamefont {Bates}, \citenamefont {Beck},
  \citenamefont {Becker}, \citenamefont {Brown}, \citenamefont {Casperson},
  \citenamefont {Conlin}, \citenamefont {Cullen}, \citenamefont {Descalle},
  \citenamefont {Firestone}, \citenamefont {Gaines}, \citenamefont {Guber},
  \citenamefont {Hawari}, \citenamefont {Holmes}, \citenamefont {Johnson},
  \citenamefont {Kawano}, \citenamefont {Kiedrowski}, \citenamefont {Koning},
  \citenamefont {Kopecky}, \citenamefont {Leal}, \citenamefont {Lestone},
  \citenamefont {Lubitz}, \citenamefont {Dami{\'a}n}, \citenamefont {Mattoon},
  \citenamefont {McCutchan}, \citenamefont {Mughabghab}, \citenamefont
  {Navratil}, \citenamefont {Neudecker}, \citenamefont {Nobre}, \citenamefont
  {Noguere}, \citenamefont {Paris}, \citenamefont {Pigni}, \citenamefont
  {Plompen}, \citenamefont {Pritychenko}, \citenamefont {Pronyaev},
  \citenamefont {Roubtsov}, \citenamefont {Rochman}, \citenamefont {Romano},
  \citenamefont {Schillebeeckx}, \citenamefont {Simakov}, \citenamefont {Sin},
  \citenamefont {Sirakov}, \citenamefont {Sleaford}, \citenamefont {Sobes},
  \citenamefont {Soukhovitskii}, \citenamefont {Stetcu}, \citenamefont {Talou},
  \citenamefont {Thompson}, \citenamefont {van~der Marck}, \citenamefont
  {Welser-Sherrill}, \citenamefont {Wiarda}, \citenamefont {White},
  \citenamefont {Wormald}, \citenamefont {Wright}, \citenamefont {Zerkle},
  \citenamefont {{\v Z}erovnik},\ and\ \citenamefont {Zhu}}]{ENDF}%
  \BibitemOpen
  \bibfield  {author} {\bibinfo {author} {\bibfnamefont {D.}~\bibnamefont
  {Brown}}, \bibinfo {author} {\bibfnamefont {M.}~\bibnamefont {Chadwick}},
  \bibinfo {author} {\bibfnamefont {R.}~\bibnamefont {Capote}}, \bibinfo
  {author} {\bibfnamefont {A.}~\bibnamefont {Kahler}},  \emph {et~al.},\
  }\enquote {\bibinfo {title} {ENDF/B-VIII.0: The 8th Major Release of the
  Nuclear Reaction Data Library with CIELO-project Cross Sections, New
  Standards and Thermal Scattering Data},}\ \href{\doibase
  10.1016/j.nds.2018.02.001} {\bibfield  {journal} {\bibinfo  {journal}
  {Nuclear Data Sheets}\ }\textbf {\bibinfo {volume} {148}},\ \bibinfo {pages}
  {1 } (\bibinfo {year} {2018})}\BibitemShut {NoStop}%
\bibitem [{\citenamefont {Hayes}\ \emph {et~al.}(2018)\citenamefont {Hayes},
  \citenamefont {Jungman}, \citenamefont {McCutchan}, \citenamefont {Sonzogni},
  \citenamefont {Garvey},\ and\ \citenamefont {Wang}}]{hayes-evolution}%
  \BibitemOpen
  \bibfield  {author} {\bibinfo {author} {\bibfnamefont {A.~C.}\ \bibnamefont
  {Hayes}}, \bibinfo {author} {\bibfnamefont {G.}~\bibnamefont {Jungman}},
  \bibinfo {author} {\bibfnamefont {E.~A.}\ \bibnamefont {McCutchan}}, \bibinfo
  {author} {\bibfnamefont {A.~A.}\ \bibnamefont {Sonzogni}}, \bibinfo {author}
  {\bibfnamefont {G.~T.}\ \bibnamefont {Garvey}}, \ and\ \bibinfo {author}
  {\bibfnamefont {X.~B.}\ \bibnamefont {Wang}},\ }\enquote {\bibinfo {title}
  {Analysis of the Daya Bay Reactor Antineutrino Flux Changes with Fuel
  Burnup},}\ \href{\doibase 10.1103/PhysRevLett.120.022503} {\bibfield
  {journal} {\bibinfo  {journal} {Phys. Rev. Lett.}\ }\textbf {\bibinfo
  {volume} {120}},\ \bibinfo {pages} {022503} (\bibinfo {year}
  {2018})}\BibitemShut {NoStop}%
\bibitem [{\citenamefont {An}\ \emph {et~al.}(2017)\citenamefont {An},
  \citenamefont {Balantekin}, \citenamefont {Band}, \citenamefont {Bishai},
  \citenamefont {Blyth}, \citenamefont {Cao}, \citenamefont {Cao},
  \citenamefont {Cao}, \citenamefont {Chan}, \citenamefont {Chang},
  \citenamefont {Chang}, \citenamefont {Chen}, \citenamefont {Chen},
  \citenamefont {Chen}, \citenamefont {Chen}, \citenamefont {Chen},
  \citenamefont {Cheng}, \citenamefont {Cheng}, \citenamefont {Cherwinka},
  \citenamefont {Chu}, \citenamefont {Chukanov}, \citenamefont {Cummings},
  \citenamefont {Ding}, \citenamefont {Diwan}, \citenamefont {Dolgareva},
  \citenamefont {Dove}, \citenamefont {Dwyer}, \citenamefont {Edwards},
  \citenamefont {Gill}, \citenamefont {Gonchar}, \citenamefont {Gong},
  \citenamefont {Gong}, \citenamefont {Grassi}, \citenamefont {Gu},
  \citenamefont {Guo}, \citenamefont {Guo}, \citenamefont {Guo}, \citenamefont
  {Guo}, \citenamefont {Hackenburg}, \citenamefont {Hans}, \citenamefont {He},
  \citenamefont {Heeger}, \citenamefont {Heng}, \citenamefont {Higuera},
  \citenamefont {Hsiung}, \citenamefont {Hu}, \citenamefont {Hu}, \citenamefont
  {Huang}, \citenamefont {Huang}, \citenamefont {Huang}, \citenamefont {Huang},
  \citenamefont {Huber}, \citenamefont {Huo}, \citenamefont {Hussain},
  \citenamefont {Jaffe}, \citenamefont {Jen}, \citenamefont {Ji}, \citenamefont
  {Ji}, \citenamefont {Jiao}, \citenamefont {Johnson}, \citenamefont {Jones},
  \citenamefont {Kang}, \citenamefont {Kettell}, \citenamefont {Khan},
  \citenamefont {Kohn}, \citenamefont {Kramer}, \citenamefont {Kwan},
  \citenamefont {Kwok}, \citenamefont {Langford}, \citenamefont {Lau},
  \citenamefont {Lebanowski}, \citenamefont {Lee}, \citenamefont {Lee},
  \citenamefont {Lei}, \citenamefont {Leitner}, \citenamefont {Leung},
  \citenamefont {Li}, \citenamefont {Li}, \citenamefont {Li}, \citenamefont
  {Li}, \citenamefont {Li}, \citenamefont {Li}, \citenamefont {Li},
  \citenamefont {Li}, \citenamefont {Li}, \citenamefont {Li}, \citenamefont
  {Li}, \citenamefont {Li}, \citenamefont {Liang}, \citenamefont {Lin},
  \citenamefont {Lin}, \citenamefont {Lin}, \citenamefont {Lin}, \citenamefont
  {Lin}, \citenamefont {Ling}, \citenamefont {Link}, \citenamefont
  {Littenberg}, \citenamefont {Littlejohn}, \citenamefont {Liu}, \citenamefont
  {Liu}, \citenamefont {Loh}, \citenamefont {Lu}, \citenamefont {Lu},
  \citenamefont {Lu}, \citenamefont {Luk}, \citenamefont {Ma}, \citenamefont
  {Ma}, \citenamefont {Ma}, \citenamefont {Malyshkin}, \citenamefont
  {Martinez~Caicedo}, \citenamefont {McDonald}, \citenamefont {McKeown},
  \citenamefont {Mitchell}, \citenamefont {Nakajima}, \citenamefont
  {Napolitano}, \citenamefont {Naumov}, \citenamefont {Naumova}, \citenamefont
  {Ngai}, \citenamefont {Ochoa-Ricoux}, \citenamefont {Olshevskiy},
  \citenamefont {Pan}, \citenamefont {Park}, \citenamefont {Patton},
  \citenamefont {Pec}, \citenamefont {Peng}, \citenamefont {Pinsky},
  \citenamefont {Pun}, \citenamefont {Qi}, \citenamefont {Qi}, \citenamefont
  {Qian}, \citenamefont {Qiu}, \citenamefont {Raper}, \citenamefont {Ren},
  \citenamefont {Rosero}, \citenamefont {Roskovec}, \citenamefont {Ruan},
  \citenamefont {Steiner}, \citenamefont {Stoler}, \citenamefont {Sun},
  \citenamefont {Tang}, \citenamefont {Taychenachev}, \citenamefont {Treskov},
  \citenamefont {Tsang}, \citenamefont {Tull}, \citenamefont {Viaux},
  \citenamefont {Viren}, \citenamefont {Vorobel}, \citenamefont {Wang},
  \citenamefont {Wang}, \citenamefont {Wang}, \citenamefont {Wang},
  \citenamefont {Wang}, \citenamefont {Wang}, \citenamefont {Wang},
  \citenamefont {Wang}, \citenamefont {Wang}, \citenamefont {Wang},
  \citenamefont {Wei}, \citenamefont {Wen}, \citenamefont {Whisnant},
  \citenamefont {White}, \citenamefont {Whitehead}, \citenamefont {Wise},
  \citenamefont {Wong}, \citenamefont {Wong}, \citenamefont {Worcester},
  \citenamefont {Wu}, \citenamefont {Wu}, \citenamefont {Wu}, \citenamefont
  {Xia}, \citenamefont {Xia}, \citenamefont {Xing}, \citenamefont {Xu},
  \citenamefont {Xu}, \citenamefont {Xue}, \citenamefont {Yang}, \citenamefont
  {Yang}, \citenamefont {Yang}, \citenamefont {Yang}, \citenamefont {Yang},
  \citenamefont {Yang}, \citenamefont {Ye}, \citenamefont {Ye}, \citenamefont
  {Yeh}, \citenamefont {Young}, \citenamefont {Yu}, \citenamefont {Zeng},
  \citenamefont {Zhan}, \citenamefont {Zhang}, \citenamefont {Zhang},
  \citenamefont {Zhang}, \citenamefont {Zhang}, \citenamefont {Zhang},
  \citenamefont {Zhang}, \citenamefont {Zhang}, \citenamefont {Zhang},
  \citenamefont {Zhang}, \citenamefont {Zhang}, \citenamefont {Zhang},
  \citenamefont {Zhang}, \citenamefont {Zhang}, \citenamefont {Zhao},
  \citenamefont {Zhou}, \citenamefont {Zhuang},\ and\ \citenamefont
  {Zou}}]{DayaBay-new}%
  \BibitemOpen
  \bibfield  {author} {\bibinfo {author} {\bibfnamefont {F.~P.}\ \bibnamefont
  {An}}, \bibinfo {author} {\bibfnamefont {A.~B.}\ \bibnamefont {Balantekin}},
  \bibinfo {author} {\bibfnamefont {H.~R.}\ \bibnamefont {Band}}, \bibinfo
  {author} {\bibfnamefont {M.}~\bibnamefont {Bishai}},  \emph {et~al.}
  (\bibinfo {collaboration} {Daya Bay Collaboration}),\ }\enquote {\bibinfo
  {title} {Evolution of the Reactor Antineutrino Flux and Spectrum at Daya
  Bay},}\ \href{\doibase 10.1103/PhysRevLett.118.251801} {\bibfield  {journal}
  {\bibinfo  {journal} {Phys. Rev. Lett.}\ }\textbf {\bibinfo {volume} {118}},\
  \bibinfo {pages} {251801} (\bibinfo {year} {2017})}\BibitemShut {NoStop}%
\bibitem [{\citenamefont {Learned}\ \emph {et~al.}(2008)\citenamefont
  {Learned}, \citenamefont {Dye}, \citenamefont {Pakvasa},\ and\ \citenamefont
  {Svoboda}}]{learned}%
  \BibitemOpen
  \bibfield  {author} {\bibinfo {author} {\bibfnamefont {J.~G.}\ \bibnamefont
  {Learned}}, \bibinfo {author} {\bibfnamefont {S.~T.}\ \bibnamefont {Dye}},
  \bibinfo {author} {\bibfnamefont {S.}~\bibnamefont {Pakvasa}}, \ and\
  \bibinfo {author} {\bibfnamefont {R.~C.}\ \bibnamefont {Svoboda}},\ }\enquote
  {\bibinfo {title} {Determination of neutrino mass hierarchy and
  ${\ensuremath{\theta}}_{13}$ with a remote detector of reactor
  antineutrinos},}\ \href{\doibase 10.1103/PhysRevD.78.071302} {\bibfield
  {journal} {\bibinfo  {journal} {Phys. Rev. D}\ }\textbf {\bibinfo {volume}
  {78}},\ \bibinfo {pages} {071302} (\bibinfo {year} {2008})}\BibitemShut
  {NoStop}%
\bibitem [{\citenamefont {Zhan}\ \emph {et~al.}(2008)\citenamefont {Zhan},
  \citenamefont {Wang}, \citenamefont {Cao},\ and\ \citenamefont {Wen}}]{zhan}%
  \BibitemOpen
  \bibfield  {author} {\bibinfo {author} {\bibfnamefont {L.}~\bibnamefont
  {Zhan}}, \bibinfo {author} {\bibfnamefont {Y.}~\bibnamefont {Wang}}, \bibinfo
  {author} {\bibfnamefont {J.}~\bibnamefont {Cao}}, \ and\ \bibinfo {author}
  {\bibfnamefont {L.}~\bibnamefont {Wen}},\ }\enquote {\bibinfo {title}
  {Determination of the neutrino mass hierarchy at an intermediate baseline},}\
  \href{\doibase 10.1103/PhysRevD.78.111103} {\bibfield  {journal} {\bibinfo
  {journal} {Phys. Rev. D}\ }\textbf {\bibinfo {volume} {78}},\ \bibinfo
  {pages} {111103} (\bibinfo {year} {2008})}\BibitemShut {NoStop}%
\bibitem [{\citenamefont {Hayes}\ and\ \citenamefont
  {Vogel}(2016)}]{hayes-vogel}%
  \BibitemOpen
  \bibfield  {author} {\bibinfo {author} {\bibfnamefont {A.~C.}\ \bibnamefont
  {Hayes}}\ and\ \bibinfo {author} {\bibfnamefont {P.}~\bibnamefont {Vogel}},\
  }\enquote {\bibinfo {title} {Reactor Neutrino Spectra},}\ \href{\doibase
  10.1146/annurev-nucl-102115-044826} {\bibfield  {journal} {\bibinfo
  {journal} {Annual Review of Nuclear and Particle Science}\ }\textbf {\bibinfo
  {volume} {66}},\ \bibinfo {pages} {219} (\bibinfo {year} {2016})}\BibitemShut
  {NoStop}%
\bibitem [{\citenamefont {Grassi}(2017)}]{grassi}%
  \BibitemOpen
  \bibfield  {author} {\bibinfo {author} {\bibfnamefont {M.}~\bibnamefont
  {Grassi}},\ }\enquote {\bibinfo {title} {The Jiangmen Underground Neutrino
  Observatory},}\ \href{\doibase 10.22323/1.274.0073} {\bibfield  {journal}
  {\bibinfo  {journal} {Proceedings of Science, XIII International Conference
  on Heavy Quarks and Leptons}\ ,\ \bibinfo {pages} {2}} (\bibinfo {year}
  {2017})}\BibitemShut {NoStop}%
\bibitem [{\citenamefont {Fornies-Marquina}\ \emph {et~al.}(1997)\citenamefont
  {Fornies-Marquina}, \citenamefont {Letosa}, \citenamefont {Garcia-Gracia},\
  and\ \citenamefont {Artacho}}]{fourierError}%
  \BibitemOpen
  \bibfield  {author} {\bibinfo {author} {\bibfnamefont {J.~M.}\ \bibnamefont
  {Fornies-Marquina}}, \bibinfo {author} {\bibfnamefont {J.}~\bibnamefont
  {Letosa}}, \bibinfo {author} {\bibfnamefont {M.}~\bibnamefont
  {Garcia-Gracia}}, \ and\ \bibinfo {author} {\bibfnamefont {J.~M.}\
  \bibnamefont {Artacho}},\ }\enquote {\bibinfo {title} {Error propagation for
  the transformation of time domain into frequency domain},}\ \href{\doibase
  10.1109/20.582534} {\bibfield  {journal} {\bibinfo  {journal} {IEEE
  Transactions on Magnetics}\ }\textbf {\bibinfo {volume} {33}},\ \bibinfo
  {pages} {1456} (\bibinfo {year} {1997})}\BibitemShut {NoStop}%
\end{thebibliography}%
\end{document}